\documentclass[aps,secnumarabic,nobalancelastpage,
nofootinbib,onecolumn,12pt]{revtex4}%
\usepackage{amsfonts}
\usepackage{amsmath}
\usepackage{graphics}
\usepackage{graphicx}
\usepackage{longtable}
\usepackage{url}
\usepackage{bm}
\usepackage{amssymb}%
\begin{document}
\title{Molecular Mean-Field Theory of Ionic Solutions: a
Poisson-Nernst-Planck-Bikerman Model}
\author{Jinn-Liang Liu}
\affiliation{Institute of Computational and Modeling Science, National Tsing Hua
University, Hsinchu 300, Taiwan; jinnliu@mail.nd.nthu.edu.tw}
\author{Bob Eisenberg}
\affiliation{Department of Physiology and Biophysics, Rush University, Chicago IL 60612
USA; beisenbe@rush.edu; Department of Applied Mathematics, Illinois Institute
of Technology, Chicago IL 60616 USA; Reisenberg@iit.edu}
\maketitle

\textbf{Abstract.} We have developed a molecular mean-field theory ---
fourth-order Poisson-Nernst-Planck-Bikerman theory --- for modeling ionic and
water flows in biological ion channels by treating ions and water molecules of
any volume and shape with interstitial voids, polarization of water, and
ion-ion and ion-water correlations. The theory can also be used to study
thermodynamic and electrokinetic properties of electrolyte solutions in
batteries, fuel cells, nanopores, porous media including cement, geothermal
brines, the oceanic system, etc. The theory can compute electric and steric
energies from all atoms in a protein and all ions and water molecules in a
channel pore while keeping electrolyte solutions in the extra- and
intracellular baths as a continuum dielectric medium with complex properties
that mimic experimental data. The theory has been verified with experiments
and molecular dynamics data from the gramicidin A channel, L-type calcium
channel, potassium channel, and sodium/calcium exchanger with real structures
from the Protein Data Bank. It was also verified with the experimental or
Monte Carlo data of electric double-layer differential capacitance and ion
activities in aqueous electrolyte solutions. We give an in-depth review of the
literature about the most novel properties of the theory, namely, Fermi
distributions of water and ions as classical particles with excluded volumes
and dynamic correlations that depend on salt concentration, composition,
temperature, pressure, far-field boundary conditions etc. in a complex and
complicated way as reported in a wide range of experiments. The dynamic
correlations are self-consistent output functions from a fourth-order
differential operator that describes ion-ion and ion-water correlations, the
dielectric response (permittivity) of ionic solutions, and the polarization of
water molecules with a single correlation length parameter.

\section{Introduction}

Water and ions give life. Their electrostatic and kinetic interactions play
essential roles in biological and chemical systems such as DNA, proteins, ion
channels, cell membranes, physiology, nanopores, supercapacitors, lithium
dendrite growth, porous media, corrosion, geothermal brines, environmental
applications, and the oceanic system
\cite{RS59,ZC86,SH90,N91,HN95,P95,A96,E96,H01,FB02,LM03,F04,LJ08,BK09,KF09,K10,E11,HR11,V11,MR12,WZ12,FK14,KS14,MT14,TE14,PD15,PW15,SO15,ZS15,ZT15,SL16,VW16,FA19,LK19}%
. Poisson, Boltzmann, Nernst, and Planck laid the foundations of classical
electrostatic and kinetic theories of ions in 1813-1890
\cite{C11,BB49,C88,N89,P90}. Gouy \cite{G10} and Chapman \cite{C13} formulated
the Poisson-Boltzmann (PB) equation in 1910 and 1913, respectively \cite{H01}.
Bikerman proposed a modified PB equation in 1942 for binary ionic liquids to
account for \textbf{different-sized ions} with \textbf{voids} \cite{B42}.
Eisenberg puns PNP for Poisson-Nernst-Planck and Positive-Negative-Positive
semiconductor transistors to emphasize nonequilibrium flows of ions through
ion channels as life's transistors \cite{EC93}. Ions in classical PB and PNP
theories are treated as volumeless point charges like the `ions' of
semiconductors, namely holes and electrons in semiconductor electronics
\cite{S50,V50,B70,K76,S76,T76,S84,JL89,MR90,J95,FG09,VG10}. Water molecules
are treated as a dielectric medium (constant) without volumes either. However,
advanced technologies in ion channel experiments \cite{SN95,B00} and material
science \cite{FA19,LK19,GM09} have raised many challenges for classical
continuum theories to describe molecular mechanisms of ions and water (or
solvents) with specific \textbf{size effects} in these systems at nano or
atomic scale \cite{H01,F04,BK09,K10,E11,HR11,V11,ZT15,FA19}.

There is another important property that classical continuum theories fail to
describe, namely, short-range ion-ion or ion-water \textbf{correlations} in
ion channels \cite{E96,H01}, charge-induced thickening and density
oscillations near highly charged surfaces \cite{BK09}, correlation-induced
charge inversion on macroions (DNA, actin, lipid membranes, colloidal
particles) \cite{S99}, the phase structure of plasma and polar fluids
\cite{L02}, colloidal charge renormalization \cite{L02}, etc. A number of
other properties related to correlations such as the dielectric response of
electrolytes solutions and the polarization of water in various conditions or
external fields are usually modeled differently from the correlation
perspective \cite{AA07,LA12,GP16}.

We have recently developed a \textbf{molecular mean-field} theory called ---
Poisson-Nernst-Planck-Bikerman (PNPB) theory --- that can describe the
\textbf{size}, \textbf{correlation}, \textbf{dielectric}, and
\textbf{polarization} effects of ions and water in aqueous electrolytes at
equilibrium or non-equilibrium all within a \textbf{unified framework}
\cite{L13,LE13,LE14,LE14a,LE15,LE15a,LH16,XL16,LX17,LE18,CC18,LL19,LL20}.
Water and ions in this theory can have \textbf{different shapes} and
\textbf{volumes} necessarily with intermolecular \textbf{voids}. The theory
generalizes and unifies the second-order Poisson-Bikerman equation \cite{B42}
of binary ionic liquids for different-sized ions having identical steric
energies \cite{LX17} and the \textbf{fourth-order differential permittivity
operator} in Santangelo's model of one component plasma \cite{S06} or in the
Bazant, Storey, and Kornyshev theory of general nonlocal permittivity for
equal-sized ions in ionic liquids \cite{BS11}.

Ion-ion and ion-water correlations are modeled by the permittivity operator
with a correlation length that depends on the diameter of ions or water and
the valence of ions of interest \cite{BS11}. The fourth-order operator yields
a permittivity as an \textbf{output function} of spatial variables, salt
concentration, and hydration shell structure including water diameter from
solving the PNPB model and thus describes the \textbf{dehydration} of ions
from bath to channel pore or from bulk to charged wall, the polarization of
water, and the change of permittivities of electrolyte solutions at different
locations in response to different configurations and conditions. Water
densities also change with configurations and conditions.

The fourth-order operator introduces correlations into the mean-field
equations so they can deal more realistically with real systems in which the
correlations are of the greatest importance. A remark should be made here that
simulations containing only particles do not automatically deal with
correlations better than mean-field theories with fourth-order operators like
this. It is not at all clear that simulations widely done in biophysics
actually compute correlations well. Indeed, it is difficult to see how
simulations that use conventions to approximate the electric field, and
periodic boundary conditions to approximate macroscopic systems could deal
with correlations correctly. The dearth of direct checks of the role of
periodic boundary conditions, and of the accuracy of the conventional
treatment of electrostatics, does little to assuage these concerns.

It is important to reiterate the obvious. Our model includes water as a
molecule and depends on the hydration structure around ions. Our model uses
partial differential equations (PDEs) to describe these essentially discrete
properties of ionic solutions, and uses the physical parameters of individual
atoms and water molecules, NOT just their mean-field description. This use of
PDEs to describe inherently discrete processes is hardly new: most of
probability theory \cite{F08,KT81} and the entire theory of wave equations,
including the wave equation of the electron called the Schr\"{o}dinger
equation \cite{S26}, treat discrete processes the same way, using PDEs that
measure (in probability theory) the underlying discrete system, or represent
it exactly as the discrete solutions of a continuum PDE (e.g., the
Schr\"{o}dinger equation describing a hydrogen atom).

The most important contribution of our work is to include water as discrete
molecules by using \textbf{Fermi distributions} \cite{F26} of classical
particles having excluded volumes with interstitial voids. We show that the
treatment of water as finite size molecules requires, as a matter of
mathematics, not physics, the existence of voids. This is demonstrated by
mathematics and simple ways to compute the voids and their role are presented.
These Fermi like distributions yield \textbf{saturation} of all particles
(ions and water) even under mathematically infinite large external fields and
\textbf{mass conservation} in the region of interest such as channel pores,
which classical theories fail to describe as well. This Fermi distribution of
classical particles obeying volume exclusion is reminiscent of the Fermi
distribution of identical particles obeying the Pauli exclusion principle
\cite{P25} in quantum mechanics.

We also introduce a new concept of distance-dependent potential between
non-bonded particles for different-sized particles similar to the electric
potential for different-charged ions and name it the \textbf{steric
potential}. The void distribution function describes the van der Waals
potential \cite{V93} of paired particles \cite{H48,RC02} in the system in a
mean-field sense. The steric potential can be written as a distribution
function of voids, emphasizing the crucial role of voids in our theory. The
specific sizes of particles and the distance-dependent steric potential allow
us to calculate steric energies at the \textbf{atomic scale}. Using Coulomb's
law allows to calculate electric energies at the atomic scale as well.
Therefore, our theory applies to biological or chemical systems having
explicit atomic structures, as well as classical mean-field representations of
bulk solutions, for example. We have shown that solving the PNPB model in
different continuum and molecular domains yields \textbf{self-consistent}
electric and steric potentials in many examples of biological ion channels or
chemical systems in
\cite{L13,LE13,LE14,LE14a,LE15,LE15a,LH16,XL16,LX17,LE18,CC18,LL19,LL20}. The
theory is also \textbf{consistent} with classical theories in the sense that
its model converges to the corresponding classical one when the volume of all
particles and the correlation length tend to zero, i.e., steric and
correlation effects vanish asymptotically to classical cases.

In this review article, we explain the above bold-face terms in detail and
compare them with those of earlier theories in a precise but limited way. The
precision means that we display explicitly, to the best of our ability, the
significant differences between analogous concepts in our theory and previous
treatments. It is obviously impossible to do complete comparisons in this vast
and formidable field. No doubt we are ignorant of significant relevant papers.
We apologize to those inadvertently slighted and ask them to help us remedy
our oversight. The remaining of this article consists as follows.

Section 2.1 describes the physical meaning of Fermi distributions and the
steric potential of ions and water with excluded volumes. We also explain the
differences between Fermi and Boltzmann distributions in the context of
statistical thermodynamics.

Section 2.2 unifies Fermi distributions and correlations into the simple and
concise 4$^{\text{th}}$-order Poisson-Bikerman (4PBik) equation. The
simplicity refers to the correlation length being the only empirical parameter
in the equation. The conciseness means that the fourth-order differential
operator can describe the complex and correlated properties of ion-ion and
ion-water interactions, polarization of water, and dielectric response of
electrolytes solutions all in a single model setting.

Section 2.3 presents a Gibbs free energy functional for the 4PBik equation. We
show that minimization of the functional yields the equation and Fermi
distributions that reduce to Boltzmann distributions when the volumes of
particles vanish in limiting case. This functional is critical to explain a
major shortcoming of earlier modified PB models that cannot yield Boltzmann
distributions in the limit. These models are thus not consistent with
classical theories and may poorly estimate steric energies and other physical
properties due to their coarse approximation of size effects.

Section 2.4 generalizes the 4PBik equation to the PNPB model to describe flow
dynamics of ions and water in the system subject to external fields. The most
important feature in this section is the introduction of the steric potential
to the classical Nernst-Planck equation. Electric and steric potentials
describe the dynamic \textbf{charge/space competition} between ions and water.
We also show that the PNPB model reduces to the 4PBik equation at equilibrium.

Section 2.5 presents a generalized Debye-H\"{u}ckel theory from the 4PBik
equation for thermodynamic modeling. The theory yields an \textbf{equation of
state} that analytically models ion activities in all types of binary and
multi-component electrolyte solutions over wide ranges of concentration,
temperature, and pressure. It is also useful to study the size, correlation,
dielectric, and polarization effects in a clear comparison with those ignoring
these effects.

Section 3 discusses numerical methods for solving the PNPB model that is
highly nonlinear and complex when coupled with the electrical field generated
by protein charges in ion channels, for example. It is very challenging to
numerically solve the model with tolerable accuracy in 3D protein structures
that generate extremely large electric field, e.g., 0.1 V in 1 Angstrom, in
parts of the molecule of great biological importance where crowded charges
directly control biological function, in the same sense that a gas pedal
controls the speed of a car.

Section 4 demonstrates the usefulness of the PNPB theory for a wide range of
biological and chemical systems, where the steric and correlation effects are
of importance. We choose a few examples of these systems, namely, electric
double layers, ion activities, and biological ion channels.

Section 5 summarizes this review with some concluding remarks.

\section{Theory}

\subsection{Fermi Distributions and Steric Potential}

The total volume of an aqueous electrolyte system with $K$ species of ions in
the solvent domain $\Omega_{s}$ is
\begin{equation}
V=\sum_{i=1}^{K+1}v_{i}N_{i}+V_{K+2}, \label{2.1}%
\end{equation}
where $K+1$ and $K+2$ denote water and voids, respectively, $v_{i}$ is the
volume of each species $i$ particle, $N_{i}$ is the total number of species
$i$ particles, and $V_{K+2}$ is the total volume of all the voids \cite{LE15}.
The volume of each particle $v_{i}$ will play a central role in our analysis,
as well that the limit $v_{i}$ goes to zero. This limit defines the solution
of point particles of classical PB and PNP theory. We must include the voids
as a separate species if we treat ions and water having volumes in a model.
This necessity can be proven by mathematics (see below). It is also apparent
to any who try to compute a model of this type with molecular water, as it was
to us \cite{LE15}.

Dividing the volume equation (\ref{2.1}) in bulk conditions by $V$, we get the
bulk volume fraction of voids
\begin{equation}
\Gamma^{B}=1-\sum_{i=1}^{K+1}v_{i}C_{i}^{B}=\frac{V_{K+2}}{V}, \label{2.2}%
\end{equation}
where $C_{i}^{B}=\frac{N_{i}}{V}$ are bulk concentrations. If the system is
spatially inhomogeneous with variable electric or steric fields, as in
realistic systems, the constants $C_{i}^{B}$ then change to functions
$C_{i}(\mathbf{r})$ and so does $\Gamma^{B}$ to a void volume function
\begin{equation}
\Gamma(\mathbf{r)}=1-\sum_{i=1}^{K+1}v_{i}C_{i}(\mathbf{r}). \label{2.3}%
\end{equation}

We define the concentrations of particles (i.e., the distribution functions of
the number density) in $\Omega_{s}$ \cite{LX17} as
\begin{equation}
C_{i}(\mathbf{r})=C_{i}^{B}\exp\left(  -\beta_{i}\phi(\mathbf{r})+\frac{v_{i}%
}{v_{0}}S^{trc}(\mathbf{r})\right)  \text{, \ }S^{trc}(\mathbf{r})=\ln
\frac{\Gamma(\mathbf{r)}}{\Gamma^{B}}, \label{2.4}%
\end{equation}
where $\phi(\mathbf{r})$ is an electric potential, $S^{trc}(\mathbf{r})$ is
called a \textbf{steric potential}, $\beta_{i}=q_{i}/k_{B}T$ with $q_{i}$
being the charge on species $i$ particles and $q_{K+1}=0$, $k_{B}$ is the
Boltzmann constant, $T$ is an absolute temperature, and $v_{0}=\left(
\sum_{i=1}^{K+1}v_{i}\right)  /(K+1)$ is an average volume. The following
inequalities%
\begin{align}
C_{i}(\mathbf{r})  &  =C_{i}^{B}\exp\left(  -\beta_{i}\phi(\mathbf{r})\right)
\left[  \frac{\Gamma(\mathbf{r)}}{\Gamma^{B}}\right]  ^{v_{i}/v_{0}}%
=\alpha_{i}\left[  1-\sum_{j=1}^{K+1}v_{j}C_{j}(\mathbf{r})\right]
^{v_{i}/v_{0}}\nonumber\\
&  =\alpha_{i}\left[  1-v_{i}C_{i}(\mathbf{r})-\sum_{j=1,j\neq i}^{K+1}%
v_{j}C_{j}(\mathbf{r})\right]  ^{v_{i}/v_{0}}<\alpha_{i}\left[  1-v_{i}%
C_{i}(\mathbf{r})\right]  ^{v_{i}/v_{0}}\nonumber\\
&  \leq\alpha_{i}\left[  1-\frac{v_{i}^{2}}{v_{0}}C_{i}(\mathbf{r})\right]
\text{ if }v_{i}/v_{0}\leq1\text{, by Bernoulli's inequality,} \label{2.5}%
\end{align}%
\begin{align}
C_{i}(\mathbf{r})  &  <\alpha_{i}\left[  1-v_{i}C_{i}(\mathbf{r})\right]
^{v_{i}/v_{0}}=\alpha_{i}\left[  1-v_{i}C_{i}(\mathbf{r})\right]  ^{\gamma
}\left[  1-v_{i}C_{i}(\mathbf{r})\right]  ^{v_{i}/v_{0}-\gamma}\nonumber\\
&  <\alpha_{i}\left[  1-v_{i}C_{i}(\mathbf{r})\right]  \left[  1-\left(
v_{i}/v_{0}-\gamma\right)  v_{i}C_{i}(\mathbf{r})\right] \nonumber\\
&  <\alpha_{i}\left[  1-v_{i}C_{i}(\mathbf{r})\right]  \text{ if }v_{i}%
/v_{0}>1, \label{2.6}%
\end{align}
imply that the distributions are of Fermi-like type \cite{K07}
\begin{align}
C_{i}(\mathbf{r})  &  <\lim_{\alpha_{i}\rightarrow\infty}\frac{\alpha_{i}%
}{1+\alpha_{i}v_{i}^{2}/v_{0}}<\frac{v_{0}}{v_{i}^{2}}\text{ if }v_{i}%
/v_{0}\leq1,\label{2.7}\\
C_{i}(\mathbf{r})  &  <\lim_{\alpha_{i}\rightarrow\infty}\frac{\alpha_{i}%
}{1+\alpha_{i}v_{i}}<\frac{1}{v_{i}}\text{ if }v_{i}/v_{0}>1, \label{2.8}%
\end{align}
i.e., $C_{i}(\mathbf{r})$ cannot exceed the maximum value $1/v_{i}^{2}$ or
$1/v_{i}$ for any arbitrary (or even infinite) potential $\phi(\mathbf{r})$ at
any location $\mathbf{r}$ in the domain $\Omega_{s}$, where $i=1,\cdots,K+1$,
$\alpha_{i}=C_{i}^{B}\exp\left(  -\beta_{i}\phi(\mathbf{r})\right)  /\left(
\Gamma^{B}\right)  ^{v_{i}/v_{0}}>0$, $0<v_{i}/v_{0}-\gamma<1$, and
$\gamma\geq1$.

The classical Boltzmann distribution appears if all particles are treated as
volumeless points, i.e., $v_{i}=0$ and $\Gamma(\mathbf{r})=\Gamma^{\text{B}%
}=1$. The classical Boltzmann distribution may produce an infinite
concentration $C_{i}(\mathbf{r})\rightarrow\infty$ in crowded conditions when
$-\beta_{i}\phi(\mathbf{r})\rightarrow\infty$, close to charged surfaces for
example, which is physically impossible \cite{L13,LE13,LE14}. This is a major,
even crippling deficiency of PB theory for modeling a system with strong local
electric fields or interactions. The difficulty in the application of
classical Boltzmann distributions to saturating systems has been avoided in
the physiological literature (apparently starting with Hodgkin, Huxley, and
Katz \cite{HH49}) by redefining the Boltzmann distribution to deal with
systems that can only exist in two states. This redefinition has been vital to
physiological research and is used in hundreds of papers \cite{B00a,BV13}, but
confusion results when the physiologists' saturating two-state Boltzmann is
not kept distinct from the unsaturating Boltzmann distribution of statistical
mechanics \cite{M76}.

It should be clearly understood that as beautiful as is Hodgkin's derivation
it begs the question of what physics creates and maintains two states. Indeed,
it is not clear how one can define the word state in a usefully unique way in
a protein of enormous molecular weight with motions covering the scale from
femtoseconds to seconds.

The steric potential $S^{trc}(\mathbf{r})$ in (\ref{2.4}) first introduced in
\cite{L13} is an entropic measure of crowding or emptiness of particles at
$\mathbf{r}$. If $\phi(\mathbf{r})=0$ and $C_{i}(\mathbf{r})=C_{i}^{B}$ then
$S^{trc}(\mathbf{r})=0$. The factor $v_{i}/v_{0}$ shows that the steric energy
$\frac{-v_{i}}{v_{0}}S^{trc}(\mathbf{r})k_{B}T$ of a type $i$ particle at
$\mathbf{r}$ depends not only on the steric potential $S^{trc}(\mathbf{r})$
but also on its volume $v_{i}$ similar to the electric energy $\beta_{i}%
\phi(\mathbf{r})k_{B}T$ depending on both $\phi(\mathbf{r})$ and $q_{i}$
\cite{LX17}, and is especially relevant to determining selectivity of specific
ions by certain biological ion channels \cite{LE13,LE14,LE15,LH16,LX17}.

In this mean-field Fermi distribution, it is impossible for a volume $v_{i}$
to be completely filled with particles, i.e., it is impossible to have
$v_{i}C_{i}(\mathbf{r})=1$ (and thus $\Gamma(\mathbf{r)}=0$) since that would
make $S^{trc}(\mathbf{r})=-\infty$ and hence $C_{i}(\mathbf{r})=0$, a
contradiction. Therefore, \emph{we must include the voids as a separate
species if we treat ions and water having volumes in a model} for which
$C_{i}(\mathbf{r})<1/v_{i}$ and $\Gamma(\mathbf{r)}\neq0$ for all
$i=1,\cdots,K+1$ and $\mathbf{r\in}$ $\Omega_{s}$. This is a critical property
distinguishing our theory from others that do not consider water as a molecule
with volume and so do not have to consider voids. We shall elaborate this
property in Section 2.3.

Our theory is consistent with the classical theory of van der Waals in
molecular physics, which describes nonbond interactions between any pair of
atoms as a distance-dependent potential such as the Lennard-Jones (L-J)
potential that cannot have zero distance between the pair \cite{H48,RC02}.

The steric potential $S^{trc}(\mathbf{r})$ lumps all van der Waals potential
energies of paired particles in a mean-field sense. It is an approximation of
L-J potentials that describe local variations of L-J distances (and thus empty
voids) between any pair of particles. L-J potentials are highly oscillatory
and extremely expensive and unstable to compute numerically \cite{LE14}.
Calculations that involve L-J potentials
\cite{V67,LM08,HV08,SK10,HF12,HL12,MB12}, or even truncated versions of L-J
potentials \cite{LnE14,G18,GE18} must be extensively checked to be sure that
results do not depend on irrelevant parameters. Any description that uses L-J
potentials has a serious problem specifying the combining rule. The details of
the combining rule \emph{directly} change predictions of effects of different
ions (selectivity) and so predictions depend on the reliability of data that
determines the combining rule and its parameters.

\emph{The steric potential does not require combining rules}. Since we
consider specific sizes of ions and water with voids, the steric potential is
valid on the \textbf{atomic scale} of L-J potentials. It is also
\textbf{consistent} with that on the \textbf{macroscopic scale} of continuum
models as shown in Sections 2.5 and 4.

To our surprise during the writing of this article, we found Eq. (2) in
Bikerman's 1942 paper \cite{B42} is \textbf{exactly} the \textbf{same} as Eq.
(\ref{2.4}) for a special case of binary ionic liquids with the identical
steric energies of different-sized ions, i.e., the factor $v_{i}/v_{0}=1$ in
(\ref{2.4}). Therefore, Bikerman's concentration function is a Fermi
distribution, a generic term used in statistical mechanics. We do NOT use
exactly the Fermi distribution as Fermi derived in 1926 for identical
particles now called fermions in quantum mechanics. So it is both more precise
and historically correct to use the name "Poisson-Bikerman"\ equation for
finite-sized ions as a generalization of the Poisson-Boltzmann equation for
volumeless ions in electrochemical and bioelectric systems.

As noted by Bazant et al. in their review paper \cite{BK09}, Bikerman's paper
has been poorly cited in the literature until recently. In our intensive and
extensive study of the literature since 2013 \cite{L13}, we have never found
any paper specifically using Bikerman's formula as Eq. (\ref{2.4}), although
of course there may be an instance we have not found. We thus now change the
term "Poisson-Fermi"\ used in our earlier papers to "Poisson-Bikerman"\ in
honor of Bikerman's brilliant work. We present here mathematical as well as
physical justifications of a very general treatment of different-sized ions
and water molecules in the mean-field framework based on Bikerman's pioneer work.

\subsection{Fourth-Order Poisson-Bikerman Equation and Correlations}

Electrolytes have been treated mostly in the tradition of physical chemistry
of isolated systems that proved so remarkably successful in understanding the
properties of ideal gases in atomic detail, long before the theory of partial
differential equations, let alone numerical computing was developed. Most
applications of ionic solutions however involve systems that are not at all
isolated. Rather most practical systems include electrodes to deliver current
and control potential, and reservoirs to manipulate the concentrations and
types of ions in the solution. Indeed, all biology occurs in ionic solutions
and nearly all of biology involves large flows. It is necessary then to extend
classical approaches so they deal with external electric fields and other
boundary conditions and allow flow so the theory can give useful results that
are applicable to most actual systems.

When the electrolyte system in $\Omega_{s}$ is subject to external fields such
as applied voltages, surface charges, and concentration gradients on the
boundary $\partial\Omega_{s}$, the electric field $\mathbf{E}(\mathbf{r})$ of
the system, the displacement field $\mathbf{D}(\mathbf{r})$ of free ions, and
the polarization field $\mathbf{P}(\mathbf{r})$ of water are generated at all
$\mathbf{r}$ in $\Omega_{s}$. In Maxwell's theory \cite{J99,Z13}, these fields
form a constitutive relation%
\begin{equation}
\mathbf{D}(\mathbf{r})=\epsilon_{0}\mathbf{E}(\mathbf{r})+\mathbf{P}%
(\mathbf{r}) \label{3.1}%
\end{equation}
and the displacement field satisfies
\begin{equation}
\nabla\cdot\mathbf{D}(\mathbf{r})=\rho_{ion}(\mathbf{r})=\sum_{i=1}^{K}%
q_{i}C_{i}(\mathbf{r}), \label{3.2}%
\end{equation}
where $\epsilon_{0}$ is the vacuum permittivity, $\rho_{ion}(\mathbf{r})$ is
the charge density of ions, and $C_{i}(\mathbf{r})$ are the concentrations
defined in (\ref{2.4}). See \cite{E19} for a modern formulation of Maxwell's
theory applicable wherever the Bohm version of quantum mechanics applies
\cite{L15,EO17}.

The electric field $\mathbf{E}(\mathbf{r})$ is thus screened by water (Bjerrum
screening) and ions (Debye screening) in a correlated manner that is usually
characterized by a correlation length $l_{c}$ \cite{HB04,S06,BS11}. The
screened force between two charges in ionic solutions (at $\mathbf{r}$ and
$\mathbf{r}^{\prime}$ in $\Omega_{s}$) has been studied extensively in
classical field theory and is often described by the van der Waals potential
kernel \cite{V93,R89,HB04,XL16,LX17}
\begin{equation}
W(\mathbf{r}-\mathbf{r}^{\prime})=\frac{e^{-\left\vert \mathbf{r}%
-\mathbf{r}^{\prime}\right\vert /l_{c}}}{\left\vert \mathbf{r}-\mathbf{r}%
^{\prime}\right\vert /l_{c}} \label{3.2a}%
\end{equation}
that satisfies the Laplace-Poisson equation \cite{R89}
\begin{equation}
-\Delta W(\mathbf{r}-\mathbf{r}^{\prime})+\frac{1}{l_{c}^{2}}W(\mathbf{r}%
-\mathbf{r}^{\prime})=\delta(\mathbf{r}-\mathbf{r}^{\prime})\text{,
\ }\mathbf{r}\text{, }\mathbf{r}^{\prime}\in R^{3} \label{3.3}%
\end{equation}
in the whole space $R^{3}$, where $\Delta=\nabla\cdot\nabla=\nabla^{2}$ is the
Laplace operator with respect to $\mathbf{r}$ and $\delta(\mathbf{r}%
-\mathbf{r}^{\prime})$ is the Dirac delta function at $\mathbf{r}^{\prime}$.

The potential $\widetilde{\phi}(\mathbf{r})$ defined in
\begin{equation}
\mathbf{D}(\mathbf{r})=-\epsilon_{s}\nabla\widetilde{\phi}(\mathbf{r})
\label{3.3a}%
\end{equation}
\cite{HB04,LX17} describes an electric potential of free ions that are
correlated only by the mean electric field according to the Poisson equation%
\begin{equation}
-\epsilon_{s}\Delta\widetilde{\phi}(\mathbf{r})=\rho_{ion}(\mathbf{r}),
\label{3.4}%
\end{equation}
a second-order partial differential equation, where $\epsilon_{s}=\epsilon
_{w}\epsilon_{0}$ and $\epsilon_{w}$ is the dielectric constant of water. This
potential does not account for correlation energies between individual ions or
between ion and polarized water in high field or crowded conditions under
which the size and valence of ions and the polarization of water play
significant roles \cite{HB04,S06,E11,BS11,LE14,LE13,LE14a,LE15}.

The correlations implicit in Maxwell's equations are of the mean-field and can
be summarized by the statement that current is conserved perfectly and
universally on all scales that the Maxwell equations are valid, where current
includes the term $\epsilon_{0}\frac{\partial\mathbf{E}(\mathbf{r}%
,t)}{\partial t}$. This term allows the Maxwell equations to describe the
propagation of light through a vacuum, and it allows charge to be
relativistically invariant, i.e., independent of velocity unlike mass, length,
and time all of which vary dramatically as velocities approach the speed of
light \cite{E19,EO17}.

We introduce the \textbf{correlated} electric potential
\begin{equation}
\phi(\mathbf{r})=\int_{R^{3}}\frac{1}{l_{c}^{2}}W(\mathbf{r}-\mathbf{r}%
^{\prime})\widetilde{\phi}(\mathbf{r}^{\prime})d\mathbf{r}^{\prime}
\label{3.5}%
\end{equation}
in \cite{LX17} as a convolution of the displacement potential $\widetilde
{\phi}(\mathbf{r}^{\prime})$ with $W(\mathbf{r}-\mathbf{r}^{\prime})$ to deal
with the correlation and polarization effects in electrolyte solutions.
However, it would be too expensive to calculate $\phi(\mathbf{r})$ using
(\ref{3.5}). Multiplying (\ref{3.3}) by $\widetilde{\phi}(\mathbf{r}^{\prime
})$ and then integrating over $R^{3}$ with respect to $\mathbf{r}^{\prime}$
\cite{XL16}, we obtain
\begin{equation}
-l_{c}^{2}\Delta\phi(\mathbf{r})+\phi(\mathbf{r})=\widetilde{\phi}(\mathbf{r})
\label{3.6}%
\end{equation}
a Laplace-Poisson equation \cite{R89,HB04} that satisfies (\ref{3.5}) in the
whole unbounded space $R^{3}$ with the boundary conditions $\phi
(\mathbf{r})=\widetilde{\phi}(\mathbf{r})=0$ at infinity. From (\ref{3.4}) and
(\ref{3.6}), we obtain the \textbf{4}$^{\text{\textbf{th}}}$\textbf{-order}
Poisson-Bikerman equation
\begin{equation}
\epsilon_{s}\left[  l_{c}^{2}\Delta-1\right]  \Delta\phi(\mathbf{r}%
)=\rho_{ion}(\mathbf{r}),\text{\ }\mathbf{r}\in\Omega_{s}, \label{3.7}%
\end{equation}
a PDE that is an approximation of (\ref{3.6}) in a bounded domain $\Omega
_{s}\subset$ $R^{3}$ with suitable boundary conditions (see below) of
$\phi(\mathbf{r})$ on $\partial\Omega_{s}$. We can thus use (\ref{3.1}) to
find the polarization field%
\begin{equation}
\mathbf{P}(\mathbf{r})=\epsilon_{s}l_{c}^{2}\nabla(\Delta\phi(\mathbf{r}%
))-(\epsilon_{w}-1)\epsilon_{0}\nabla\phi(\mathbf{r}) \label{3.8}%
\end{equation}
with $\mathbf{E}(\mathbf{r})=-\nabla\phi(\mathbf{r})$. If $l_{c}=0$, we
recover the standard Poisson equation (\ref{3.4}) and the standard
polarization $\mathbf{P}=\epsilon_{0}(\epsilon_{w}-1)\mathbf{E}$ with the
electric susceptibility $\epsilon_{w}-1$ (and thus the dielectric constant
$\epsilon_{w}$) if water is treated as a time independent, isotropic, and
linear dielectric medium \cite{Z13}. In this case, the field relation
$\mathbf{D}=\epsilon_{w}\epsilon_{0}\mathbf{E}$ with the scalar constant
permittivity $\epsilon_{s}\epsilon_{0}$ is an approximation of the exact
relation (\ref{3.1}) due to the simplification of the dielectric responses of
the medium material to the electric field $\mathbf{E}$ \cite{E19a,BB95,BB01}.

The exponential van der Waals potential $W(\mathbf{r}-\mathbf{r}^{\prime
})=\frac{e^{-\left\vert \mathbf{r}-\mathbf{r}^{\prime}\right\vert /l_{c}}%
}{\left\vert \mathbf{r}-\mathbf{r}^{\prime}\right\vert /l_{c}}$ \cite{V93} is
called the Yukawa \cite{Y35} potential in \cite{XL16,LX17} and usually in
physics, which is an anachronism \cite{R89,R05}. Van der Waals derived this
potential in his theory of capillarity based on the proposition that the
intermolecular potential of liquids and gases is shorter-ranged, but much
stronger than Coulomb's electric potential \cite{R89}. Ornstein and Zernike
(OZ) introduced short- (direct) and long-ranged (indirect) correlation
functions in their critical point theory \cite{OZ14}. There are three
important properties of the van der Waals potential: (i) it satisfies the
Laplace-Poisson equation (\ref{3.3}), (ii) it generates the same functional
form for short- and long-ranged correlations in the OZ theory, and (iii) it
solves van der Waals's problem for the intermolecular potential \cite{R89}.

Therefore, the potential $\phi(\mathbf{r})$ in (\ref{3.5}) includes
\textbf{correlation} energies of \textbf{ion-ion} and \textbf{ion-water}
interactions in \textbf{short} as well as \textbf{long} ranges in our system.
The \textbf{correlation length} $l_{c}$ can be derived from the OZ equation,
see Eq. (13) in \cite{R89}, but the derivation is not very useful. The
correlation length becomes an unknown functional of $\rho_{ion}(\mathbf{r})$
in (\ref{3.2}) and the OZ direct correlation function, and is hence usually
chosen as an empirical parameter to fit experimental, molecular dynamics (MD),
or Monte Carlo (MC) data
\cite{BK09,L13,LE13,LE14,LE14a,LE15,LE15a,LH16,XL16,LX17,LE18,LL19,S06,BS11,HB04}%
. It seems clear that it would be useful to have a theory that showed the
dependence of correlation length on ion composition and concentration, and
other parameters.

There are several approaches to fourth-order Poisson-Boltzmann equations for
modeling ion-ion and ion-water correlations from different perspectives of
physics \cite{S06,BS11,XL16,BM17,DB18}. In \cite{S06}, a decomposed kernel
acts on a charge density of counterions in a binary liquid without volumes and
water (ion-ion correlations) in contrast to the potential $\widetilde{\phi
}(\mathbf{r})$ in (\ref{3.5}) that is generated by different-sized ions and
water with voids in (\ref{3.4}) (ion-ion and ion-water correlations in a
multicomponent aqueous electrolyte). The kernel consists of short-range (of
van der Waals type) and long-range components from a decomposition of
Coulomb's interactions. In \cite{BS11}, the kernel is a general nonlocal
kernel that acts on a charge density of equal-sized ions in a binary liquid
without water (ion-ion correlations). The kernel is a series expansion of the
gradient operator $\nabla$ and thus can yield not only a fourth-order PB but
even higher order PDEs. The fourth-order PB is the first order approximation
of the energy expansion that converges only with small wavenumbers $k$ in
Fourier frequency domain for the dielectric response of ionic liquids
\cite{BS11}.

Derived from the framework of nonlocal electrostatics for modeling the
dielectric properties of water in \cite{HB04}, the kernel acting on
$\widetilde{\phi}(\mathbf{r})$ in \cite{XL16} (ion-ion and ion-water
correlations) consists of a van der Waals function and the Dirac delta
function that correspond to the limiting cases $k=0$ and $k=\infty$,
respectively. In \cite{BM17}, a system of three PDEs derived from
electrostatics and thermodynamic pressure has electric potential and
concentration gradients of equal-sized cations and anions in a binary fluid as
three unknown functions. Linearization and simplification of the nonlinear
system can yield a linear fourth-order PB (ion-ion correlations). In
\cite{DB18}, the fourth-order PB is derived from a free energy functional that
models ion-ion correlations in a binary liquid using volume-fraction functions
of equal-sized cations and anions with two additional parameters associated
with the interaction energies of these two functions and their gradients.

The \textbf{dielectric operator} $\epsilon_{s}\left(  l_{c}^{2}\Delta
-1\right)  $ in (\ref{3.7}) describes changes in dielectric response of water
with salt concentrations (ion-water correlations), ion-ion correlations, and
water polarizations all via the mean-field charge density function $\rho
_{ion}(\mathbf{r})$ provided that we can solve (\ref{2.4}) and (\ref{3.7}) for
a consistent potential function $\phi(\mathbf{r})$. Therefore, the operator (a
mapping) depends not only on ion and water concentrations ($C_{i}^{B}$ for all
arbitrary species $i=1,\cdots,K+1$ of particles with any arbitrary shapes and
volumes) but also on the location $\mathbf{r}$ and the voids at $\mathbf{r}$.
The operator thus produces a \textbf{dielectric function} $\widehat{\epsilon
}(\mathbf{r,}C_{i}^{B})$ as an \textbf{output} from the solution
$\phi(\mathbf{r})$ that satisfies the 4PBik equation (\ref{3.7}) that
saturates as a function of concentration (\ref{2.4}), as we shall repeatedly
emphasize. This dielectric function $\widehat{\epsilon}(\mathbf{r,}C_{i}^{B})$
is not an additional model for $\widetilde{\epsilon}(\mathbf{r})$,
$\widetilde{\epsilon}(k)$, or $\widetilde{\epsilon}(C_{i}^{B})$ as it often is
in other models in the literature
\cite{LA12,KS96,SW01,MM02,CK03,GK04,CC05,NV08,SB10,LH11,GP16,N18,K18}. Here
the dielectric function is an output, as we have stated.

The 4PBik (\ref{3.7}) with (\ref{2.4}) is a very general model using only one
extra parameter $l_{c}$ in the fourth-order operator to include many physical
properties ignored by the classical Poisson-Boltzmann equation. We shall
illustrate these properties of our model in Section 4.

\subsection{Generalized Gibbs Free Energy Functional}

To generalize the Gibbs free energy functional for Boltzmann distributions
that satisfy the classical Poisson-Boltzmann equation \cite{SH90,FB97,L09}, we
introduce a functional in \cite{LX17} for saturating Fermi distributions
(\ref{2.4}) that satisfy the 4$^{\text{th}}$-order Poisson-Bikerman equation
(\ref{3.7})%
\begin{align}
F(\mathbf{C})  &  =F_{el}(\mathbf{C})+F_{en}(\mathbf{C}),\label{4.1}\\
F_{el}(\mathbf{C})  &  =\frac{1}{2}\int_{\Omega_{s}}\rho_{ion}(\mathbf{r}%
)L^{-1}\rho_{ion}(\mathbf{r})d\mathbf{r},\text{ }\label{4.2}\\
F_{en}(\mathbf{C})  &  =k_{B}T\int_{\Omega_{s}}\left\{  \sum_{i=1}^{K+1}%
C_{i}(\mathbf{r})\left(  \ln\frac{C_{i}(\mathbf{r})}{C_{i}^{B}}-1\right)
+\frac{\Gamma(\mathbf{r})}{v_{0}}\left(  \ln\frac{\Gamma(\mathbf{r})}%
{\Gamma^{B}}-1\right)  \right\}  d\mathbf{r}, \label{4.3}%
\end{align}
where $F_{el}(\mathbf{C})$ is an electrostatic functional, $F_{en}%
(\mathbf{C})$ is an entropy functional, $\mathbf{C}=\left(  C_{1}%
(\mathbf{r})\text{, }C_{2}(\mathbf{r})\text{,}\cdots\text{, }C_{K+1}%
(\mathbf{r}\mathbb{)}\right)  $, and $L^{-1}$ is the inverse of the
self-adjoint positive linear operator $L=\epsilon_{s}\left(  l_{c}^{2}%
\Delta-1\right)  \Delta$ \cite{XL16} in (\ref{3.7}), i.e., $L\phi
(\mathbf{r})=\rho_{ion}(\mathbf{r})$. $\mathbf{C}$ is a `concentration vector'
that specifies the number density, i.e., concentration of each species in the
ionic solution, including water. $\mathbf{C}$ plays a central role in any
theory of ionic solutions because it specifies the main property of a
solution, namely its composition.

Taking the variations of $F(\mathbf{C})$ at $C_{i}(\mathbf{r})$, we have%
\begin{gather*}
\frac{\delta F(\mathbf{C})}{\delta C_{i}}=\int_{\Omega_{s}}\left\{
k_{B}T\left[  \ln\frac{C_{i}(\mathbf{r})}{C_{i}^{B}}-\frac{v_{i}}{v_{0}}%
\ln\frac{\Gamma(\mathbf{r})}{\Gamma^{B}}\right]  +\frac{1}{2}\left(
q_{i}L^{-1}\rho_{ion}(\mathbf{r})+\rho_{ion}(\mathbf{r})L^{-1}q_{i}\right)
\right\}  d\mathbf{r,}\\
\frac{1}{2}\left(  q_{i}L^{-1}\rho_{ion}(\mathbf{r})+\rho_{ion}(\mathbf{r}%
)L^{-1}q_{i}\right)  =q_{i}\phi(\mathbf{r}),
\end{gather*}%
\begin{equation}
\frac{\delta F(\mathbf{C})}{\delta C_{i}}=0\Rightarrow k_{B}T\left[  \ln
\frac{C_{i}(\mathbf{r})}{C_{i}^{B}}-\frac{v_{i}}{v_{0}}\ln\frac{\Gamma
(\mathbf{r})}{\Gamma^{B}}\right]  +q_{i}\phi(\mathbf{r})=0 \label{4.4}%
\end{equation}
that yields the saturating Fermi distributions in (\ref{2.4}) for all
$i=1,\cdots,K+1$. Moreover, we have%
\begin{equation}
\frac{\delta^{2}F(\mathbf{C})}{\delta C_{i}^{2}}=\int_{\Omega_{s}}\left\{
k_{B}T\left[  \frac{1}{C_{i}(\mathbf{r})}+\frac{v_{i}^{2}}{v_{0}}\frac
{\Gamma^{B}}{\Gamma(\mathbf{r})}\right]  +q_{i}^{2}L^{-1}C_{i}\right\}
d\mathbf{r}>0 \label{4.5}%
\end{equation}
implying that the saturating Fermi distribution vector $\mathbf{C}$ is a
unique minimizer of the functional $F(\mathbf{C})$.

The Gibbs-Bikerman free energy functional $F(\mathbf{C})$ has two important
properties. First, its electrostatic part $F_{el}(\mathbf{C})$ is defined in
terms of the composition vector $\mathbf{C}$ only. It depends only on
concentrations and nothing else. If an electrostatic functional $\widetilde
{F}_{el}(\widetilde{\phi}(\mathbf{r}))$ is defined in terms of $\left\vert
\nabla\widetilde{\phi}(\mathbf{r})\right\vert ^{2}$ for the PB equation
\cite{L13,BS11,SH90a,RR90,GD93,BA97,SB10,BA11,LZ11,ZW11,QT14}, the
corresponding concentration vector $\widetilde{\mathbf{C}}$ and the potential
$\widetilde{\phi}(\mathbf{r})$ do \emph{not} minimize the corresponding
functional $\widetilde{F}(\widetilde{\mathbf{C}},\widetilde{\phi}%
(\mathbf{r}))$ \cite{FB97,L09}, i.e., $\widetilde{F}$ is not a Gibbs free
energy functional \cite{SH90,FB97}. Second, the limit of its entropic part
\begin{equation}
\lim_{v_{i}\rightarrow0}F_{en}(\mathbf{C})=k_{B}T\int_{\Omega_{s}}\sum
_{i=1}^{K+1}C_{i}^{0}(\mathbf{r})\left(  \ln\frac{C_{i}^{0}(\mathbf{r})}%
{C_{i}^{B}}-1\right)  d\mathbf{r} \label{4.6}%
\end{equation}
exists ($F_{en}$ converges) when the volume $v_{i}$ tends to zero for all
$i=1,\cdots,K+1$. This implies that all ionic species have Boltzmann
distributions $C_{i}^{0}(\mathbf{r})=C_{i}^{B}\exp\left(  -\beta_{i}%
\phi(\mathbf{r})\right)  $, $i=1,\cdots,K$, the water density $C_{K+1}%
^{0}(\mathbf{r})=C_{K+1}^{B}$ is a constant, and the void fraction
$\Gamma(\mathbf{r})=\Gamma^{B}=1$ since all particles are volumeless in PB
theory. Therefore, the 4PBik model (\ref{2.4}) and (\ref{3.7}) is physically
and mathematically \textbf{consistent} with the classical PB model in the
limiting case when we ignore the steric ($v_{i}=0$) and correlation ($l_{c}=0
$) effects.

There are many shortcomings about the lattice approach \cite{GM47} frequently
used to account for steric effects in lattice-based PB models
\cite{BK09,AA07,L13,BS11,SB10,L09,BA97,T08,BA11,LZ11,ZW11,BK01,MG18}. For
example, (i) it assumes equal-sized ions and thus cannot distinguish
non-uniform particles as in (\ref{2.1}), (ii) its effective ion size needs to
be unrealistically large to fit data \cite{BK09}, (iii) its correction over
Boltzmann's point charge approach appears only at high surface charges
\cite{LH11}, (iv) its pressure term diverges very weakly (is greatly
underestimated) at close packing \cite{MP16}, and (v) its entropy functional
may diverge as the volume of ions tends to zero, i.e., the corresponding
\textbf{lattice-based PB} model is \textbf{not} physically and mathematically
\textbf{consistent} with the classical PB model in the limiting case
\cite{LE14}.

The importance of the restriction in Point (i) is hard to overstate. Almost
all the interesting properties of ionic solutions arise because of their
selectivity (as it is called in biology) or specificity between species, and
those different properties arise in large measure because of the different
diameters of the ions. The equal diameter case is dull and degenerate.

Point (v) is a critical problem that is closely related to Points (ii) - (iv).
The divergence is obvious for an entropy term $\widetilde{F}_{en}$ in Eq. (2)
in \cite{BA97} as%
\begin{equation}
\lim_{v\rightarrow0}\widetilde{F}_{en}=\lim_{v\rightarrow0}\sum_{i=1}%
^{K}\widetilde{C}_{i}(\mathbf{r})\ln\left(  v\widetilde{C}_{i}(\mathbf{r}%
)\right)  =-\infty, \label{4.7}%
\end{equation}
which also appears in
\cite{AA07,L13,BS11,SB10,L09,BA97,T08,BA11,LZ11,ZW11,BK01,MG18}. It is
impossible to derive Boltzmann distributions $\widetilde{C}_{i}(\mathbf{r}%
)=C_{i}^{B}\exp\left(  -\beta_{i}\widetilde{\phi}(\mathbf{r})\right)  $ from
$\widetilde{F}_{en}$ as $v\rightarrow0$ without extra assumptions, see (2.6)
in \cite{L09}, for example. In fact, the assumption (2.6), i.e.,
$v\widetilde{C}_{i}(\mathbf{r})>0$, actually forbids us from taking $v$ to the
limit zero.

Our derivation of $F_{en}(\mathbf{C})$ does not employ any lattice models but
simply uses the exact volume equation (\ref{2.1}). Our theory should not be
classified then as a lattice model as sometimes is the case, at least in
informal discussions. The void function $\Gamma(\mathbf{r)}$ is an analytical
generalization of the void fraction $1-\Phi$ in (20) in \cite{BK09} with all
volume parameters $v_{i}$ (including the bulk fraction $\Gamma^{B}$) being
physical instead of empirical as $\Phi$. The excess chemical potential in
\cite{BK09} is $-k_{B}T\ln(1-\Phi)$ whereas ours is $F_{en}(\mathbf{C})$ in
(\ref{4.3}).

These expressions are different in important respects. Our model is not a
lattice-based model because its differences are crucial both mathematically
and physically. Indeed, the lattice-based model is in a certain sense
internally inconsistent with classical statistical mechanics since a
fundamental result of classical statistical mechanics $v\widetilde{C}%
_{i}(\mathbf{r})>0$ prevents the model from satisfying the classical
imperative of the Boltzmann distribution in the limit of zero $v$.

The Langmuir-type distribution
\begin{equation}
C_{i}(x)=\frac{C_{i}^{B}\exp\left(  -\beta_{i}\phi(x)\right)  }{1+\sum
_{j=1}^{K}\frac{C_{j}^{B}}{C_{j}^{\max}}\left(  \exp\left(  -\beta_{j}%
\phi(x)\right)  -1\right)  }\text{ } \label{4.8}%
\end{equation}
of different-sized ions (without water) proposed in \cite{LH11} also reduces
to a Boltzmann distribution as $v_{j}\rightarrow0$, $\forall j$, where
$C_{j}^{\max}=p/v_{j}$ and $p\leq1$ is a packing parameter. This distribution
saturates and thus is of Fermi type, i.e., $C_{i}(x)\leq C_{i}^{\max}$ and
$v_{i}C_{i}(x)\leq1$. The entropy term $-\ln\left(  1+\sum_{j=1}^{K}%
\frac{C_{j}^{B}}{C_{j}^{\max}}\left(  \exp\left(  -\beta_{j}\phi(x)\right)
-1\right)  \right)  $ does not involve voids so is different from the
$S^{trc}(\mathbf{r})$ in (\ref{2.4}). Our distribution in (\ref{2.4}) does not
need any packing parameters and satisfies $v_{i}C_{i}(\mathbf{r})<1$.\newline

\subsection{Poisson-Nernst-Planck-Bikerman Model of Saturating Phenomena}

For nonequilibrium systems, we can also generalize the classical
Poisson-Nernst-Planck model \cite{EC93,N89,P90,CB92,EK95} to the
Poisson-Nernst-Planck-Bikerman model by coupling the flux density equation
\begin{equation}
\frac{\partial C_{i}(\mathbf{r},t)}{\partial t}=-\nabla\cdot\mathbf{J}%
_{i}(\mathbf{r},t),\text{ }\mathbf{r}\in\Omega_{s} \label{5.1}%
\end{equation}
of each particle species $i=1,\cdots,K+1$ (including water) to the 4PBik
equation (\ref{3.7}), where the flux density is defined as%
\begin{equation}
\mathbf{J}_{i}(\mathbf{r},t)=-D_{i}\left[  \nabla C_{i}(\mathbf{r}%
,t)+\beta_{i}C_{i}(\mathbf{r},t)\nabla\phi(\mathbf{r},t)-\frac{v_{i}}{v_{0}%
}C_{i}(\mathbf{r},t)\nabla S^{trc}(\mathbf{r},t)\right]  , \label{5.2}%
\end{equation}
$D_{i}$ is the diffusion coefficient, and the time variable $t$ is added to
describe the dynamics of electric $\phi(\mathbf{r},t)$ and steric
$S^{trc}(\mathbf{r},t)$ potentials.

The flux equation (\ref{5.1}) is called the Nernst-Planck-Bikerman equation
because the steric potential $S^{trc}(\mathbf{r},t)$ is introduced to the
classical NP equation so it can deal with saturating phenomena including those
that arise from the unequal volumes of ions and the finite volume of molecular
water. The PNPB model can be extended to include hydrodynamic kinetic and
potential energies in the variational treatment of energy processes (i.e.,
EnVarA) by Hamilton's least action and Rayleigh's dissipation principles
\cite{EH10,LW19}. We shall however consider this as a topic for future work.

At equilibrium, the net flow of each particle species is a zero vector, i.e.,
$\mathbf{J}_{i}(\mathbf{r})=\mathbf{0}$ (in a steady state), which implies
that%
\begin{align}
\nabla C_{i}(\mathbf{r})+\beta_{i}C_{i}(\mathbf{r})\nabla\phi(\mathbf{r}%
)-\frac{v_{i}}{v_{0}}C_{i}(\mathbf{r})\nabla S^{trc}(\mathbf{r})  &
=\mathbf{0},\nonumber\\
\nabla\left[  C_{i}(\mathbf{r})\exp(\beta_{i}\phi(\mathbf{r})-\frac{v_{i}%
}{v_{0}}S^{trc}(\mathbf{r}))\right]   &  =\mathbf{0},\nonumber\\
C_{i}(\mathbf{r})\exp(\beta_{i}\phi(\mathbf{r})-\frac{v_{i}}{v_{0}}%
S^{trc}(\mathbf{r}))  &  =c_{i}, \label{5.3}%
\end{align}
where the constant $c_{i}=C_{i}^{B}$ for $\phi(\mathbf{r})=S^{trc}%
(\mathbf{r})=0$. Therefore, (\ref{5.3}) $=$ (\ref{2.4}), i.e., the NPF
equation (\ref{5.1}) reduces to the saturating Fermi distribution (\ref{2.4})
as the classical NP equation reduces to the Boltzmann distribution at equilibrium.

The gradient of the steric potential $\nabla S^{trc}(\mathbf{r},t)$ in
(\ref{5.2}) represents an entropic force of vacancies exerted on particles.
The negative sign in $-C_{i}(\mathbf{r},t)\nabla S^{trc}(\mathbf{r},t)$ means
that the steric force $\nabla S^{trc}(\mathbf{r},t)$ is in the opposite
direction to the diffusion force $\nabla C_{i}(\mathbf{r},t)$.

Larger $S^{trc}(\mathbf{r},t)=\ln\frac{\Gamma(\mathbf{r},t)}{\Gamma^{B}}$
implies lower pressure because the ions occupy more space (less crowded) as
implied by the numerator $\Gamma(\mathbf{r},t)$. The larger the $S^{trc}%
(\mathbf{r},t)$ the lower pressure at the location $\mathbf{r}$, the more the
entropic force (the higher pressure) pushes particles to $\mathbf{r}$ from
neighboring locations. The steric force is the opposite of the diffusion force
$\nabla C_{i}(\mathbf{r},t)$ that pushes particles away from $\mathbf{r}$ if
the concentration at $\mathbf{r}$ is larger than that at neighboring locations.

Moreover, the Nernst-Einstein relationship between diffusion and mobility
\cite{H01} implies that the steric flux $D_{i}\frac{v_{i}}{v_{0}}%
C_{i}(\mathbf{r},t)\nabla S^{trc}(\mathbf{r},t)$ is greater if the particle is
more mobile. The Nernst-Einstein relationship is generalized to
\begin{equation}
\mu_{i}=v_{i}q_{i}D_{i}/(v_{0}k_{B}T), \label{5.4}%
\end{equation}
where the mobility coefficient $\mu_{i}$ of an ion depends on its size $v_{i}$
in addition to its charge $q_{i}$. The mobility coefficient of water is
$\mu_{K+1}=v_{K+1}D_{K+1}/(v_{0}k_{B}T)$.

Therefore, the gradients of electric and steric potentials ($\nabla
\phi(\mathbf{r},t)$ and $\nabla S^{trc}(\mathbf{r},t)$) describe the
\textbf{charge/space competition} mechanism of particles in a crowded region
within a mean-field framework. Since $S^{trc}(\mathbf{r},t)$ describes the
dynamics of void movements, the dynamic crowdedness (pressure) of the flow
system can also be quantified. A large amount of experimental data exists
concerning the dependence of diffusion coefficient on the concentration and
size of solutes. Comparing our model with this data is an important topic of
future work.

The motion of water molecules, i.e., the \textbf{osmosis} of water
\cite{XE18,ZX19} is directly controlled by the steric potential in our model
and their distributions are expressed by $C_{K+1}(\mathbf{r},t)=C_{K+1}%
^{B}\exp\left(  v_{K+1}S^{trc}(\mathbf{r},t)/v_{0}\right)  $. Nevertheless,
this motion is still implicitly changed by the electric potential
$\phi(\mathbf{r},t)$ via the correlated motion of ions described by other
$C_{j}(\mathbf{r},t)$ in the void fraction function $\Gamma(\mathbf{r},t)$ and
hence in the charge density $\rho_{ion}(\mathbf{r},t)$ in (\ref{3.7}).

In summary, the PNPB model accounts for (i) the \textbf{steric}
(\textbf{pressure}) effect of ions and water molecules, (ii) the
\textbf{correlation} effect of crowded ions, (iii) the \textbf{screening}
(\textbf{polarization}) effect of polar water, and (iv) the
\textbf{charge/space competition} effect of ions and water molecules of
different sizes and valences. These effects are all closely related to the
interstitial voids between particles and described by two additional terms,
namely, the \textbf{correlation length} and the \textbf{steric potential}.

\subsection{Generalized Debye-H\"{u}ckel Theory}

Thermodynamic modeling is of fundamental importance in the study of chemical
and biological systems \cite{RS59,P95,H01,LM03,F04,LJ08,K10,VW16}. Since Debye
and H\"{u}ckel (DH) proposed their theory in 1923 \cite{DH23} and H\"{u}ckel
extended it to include Born energy effects in 1925 \cite{H25}, a great variety
of extended DH models (equations of state) have been developed for modeling
aqueous or mixed-solvent solutions over wide ranges of composition,
temperature, and pressure \cite{P95,V11,MS02,VB07,RK15,KM18,BM20}. Despite
these intense efforts, robust thermodynamic modeling of electrolyte solutions
still presents a difficult challenge for extended DH models due to enormous
amount of parameters that need to be adjusted carefully and often subjectively
\cite{V11,VB07,RK15,KM18,F10}.

It is indeed a frustrating despair (the word \textit{frustration} on p. 11 in
\cite{K10} and the word \textit{despair} on p. 301 in \cite{RS59}) that about
\textbf{22,000} parameters \cite{V11} need to be extracted from the available
experimental data for one temperature for combinatorial solutions of the most
important 28 cations and 16 anions in salt chemistry by the Pitzer model
\cite{P95}, which is the most widely used DH model with unmatched precision
for modeling electrolyte solutions \cite{RK15}. The JESS (joint expert
speciation system) is the world's largest system of thermodynamic information
relating to electrolytes, reactions in aqueous media, and hydrocarbon phase
equilibria \cite{MR20}. The total number of Pitzer's fitting parameters in
JESS is \textbf{95} \cite{MR10}.

By contrast, we propose in \cite{LL19,LL20} a generalized Debye-H\"{u}ckel
theory from the 4PBik equation (\ref{3.7}) to include (i) steric effects, (ii)
correlation effects, (iii) Born solvation energy, and (iv) ion hydration
\cite{SR48,RH85,M91,OR93,BL99,VR06,MP11,RI13} that are missing in the original
DH theory. The generalized theory can be used to calculate ion activities in
all types of binary and multi-component solutions over wide ranges of
concentration, temperature, and pressure with only \textbf{3} fitting
parameters \cite{LE15a,LE18,LL19,LL20}.

We briefly outline the derivation of a generalized DH equation of state and
refer to \cite{LL20} for more details. The activity coefficient $\gamma_{i}$
of an ion of species $i$ in an aqueous electrolyte solution with a total of
$K$ species of ions describes deviation of the chemical potential of the ion
from ideality ($\gamma_{i}=1$) \cite{LM03}. The excess chemical potential
$\mu_{i}^{ex}=k_{B}T\ln\gamma_{i}$ can be calculated by \cite{LE15a,BC00}%
\begin{equation}
\mu_{i}^{ex}=\frac{1}{2}q_{i}\phi(\mathbf{0})-\frac{1}{2}q_{i}\phi
^{0}(\mathbf{0}),\label{6.1}%
\end{equation}
where $q_{i}$ is the charge of the hydrated ion (also denoted by $i$),
$\phi(\mathbf{r})$ is a reaction potential \cite{BC00} function of spatial
variable $\mathbf{r}$ in the domain $\overline{\Omega}$ $=$ $\overline{\Omega
}_{i}\cup\overline{\Omega}_{sh}\cup\overline{\Omega}_{s}$ shown in Fig. 1,
$\Omega_{i}$\ is the spherical domain occupied by the ion $i$, $\Omega_{sh}$
is the hydration shell domain of the ion, $\Omega_{s}$ is the rest of solvent
domain, $\mathbf{0}$ denotes the center (set to the origin) of the ion, and
$\phi^{0}(\mathbf{r})$ is a potential function when the solvent domain
$\Omega_{s}$ does not contain any ions at all with pure water only, i.e., when
the solution is ideal. The radii of $\Omega_{i}$ and the outer boundary of
$\Omega_{sh}$ are denoted by $R_{i}^{Born}$ (ionic cavity radius \cite{RH85})
and $R_{i}^{sh}$, respectively.   \begin{figure}[t]
\centering\includegraphics[scale=0.6]{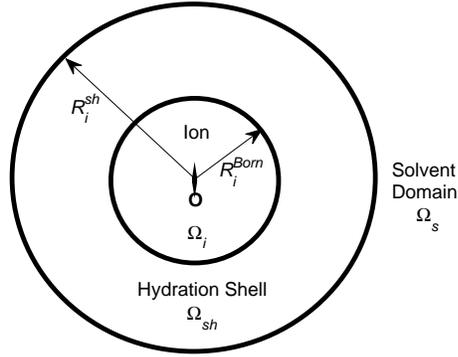}\caption{ The model domain
$\Omega$ is partitioned into the ion domain $\Omega_{i}$ (with radius
$R_{i}^{Born}$), the hydration shell domain $\Omega_{sh}$ (with radius
$R_{i}^{sh}$), and the remaining solvent domain $\Omega_{s}$.}%
\end{figure}

The potential function $\phi(\mathbf{r})$ can be found by solving the 4PBik
equation (\ref{3.7}) and the Laplace equation \cite{LE15a,LE18}%
\begin{equation}
\Delta\phi(\mathbf{r})=0\text{ in }\Omega_{i}\cup\Omega_{sh},\label{6.2}%
\end{equation}
where $\epsilon_{s}$ is defined in $\overline{\Omega}_{sh}\cup\Omega_{s}$, the
correlation length $l_{c}=\sqrt{l_{B}l_{D}/48}$ is a density-density
correlation length independent of specific ionic radius \cite{LF96}, $l_{B}$
and $l_{D}$ are the Bjerrum and Debye lengths, respectively, the concentration
$C_{k}(\mathbf{r})$ function (\ref{2.4}) is defined in $\overline{\Omega}$ for
all $k=1,\cdots,K+1$ in molarity (M), and $v_{k}=4\pi a_{k}^{3}/3$ with radius
$a_{k}$. Since the steric potential takes particle volumes and voids into
account, the shell volume $V_{sh}$ of the shell domain $\Omega_{sh}$ can be
determined by the steric potential $S_{sh}^{trc}=\frac{v_{0}}{v_{w}}\ln
\frac{O_{i}^{w}}{V_{sh}C_{K+1}^{B}}=\ln\frac{V_{sh}-v_{w}O_{i}^{w}}%
{V_{sh}\Gamma^{B}}$ \cite{LE15a,LE18}, where the occupant (coordination)
number $O_{i}^{w}$ of water molecules is given by experimental data
\cite{RI13}. The shell radius $R_{i}^{sh}$ is thus determined and depends not
only on $O_{i}^{w}$ but also on the bulk void fraction $\Gamma^{B}$, namely,
\textit{on all salt and water bulk concentrations} ($C_{k}^{B}$)
\cite{LE15a,LE18}.

For reducing the complexity of higher-order approximations in implementation,
we transform the fourth-order PDE (\ref{3.7}) to the following two
second-order PDEs \cite{L13}%
\begin{align}
\left(  l_{c}^{2}\Delta-1\right)  \psi(\mathbf{r}) &  =\rho_{ion}%
(\mathbf{r})\text{ in }\Omega_{s},\label{6.3}\\
\epsilon_{s}\Delta\phi(\mathbf{r}) &  =\psi(\mathbf{r})\text{ in }\Omega
_{s}\text{,}\label{6.4}%
\end{align}
where the extra unknown function $\psi(\mathbf{r})$ is a density-like function
as seen from (\ref{6.3}) by setting $l_{c}=0$. The boundary and interface
conditions for $\phi(\mathbf{r})$ and $\psi(\mathbf{r})$ in (\ref{6.2}%
)-(\ref{6.4}) are \cite{L13}%
\begin{align}
\phi(\mathbf{r}) &  =\psi(\mathbf{r})=0\text{ on }\partial\Omega_{s}%
\backslash\partial\Omega_{sh},\label{6.5}\\
\psi(\mathbf{r}) &  =-\rho_{s}(\mathbf{r})\text{ on }\partial\Omega_{sh}%
\cap\partial\Omega_{s},\label{6.6}\\
\left[  \phi(\mathbf{r})\right]   &  =0\text{ on }\partial\Omega_{i}%
\cup\left(  \partial\Omega_{sh}\cap\partial\Omega_{s}\right)  ,\label{6.7}\\
\left[  \nabla\phi(\mathbf{r})\cdot\mathbf{n}\right]   &  =0\text{ on
}\partial\Omega_{sh}\cap\partial\Omega_{s},\label{6.8}\\
\left[  \epsilon(\mathbf{r})\nabla\phi(\mathbf{r})\cdot\mathbf{n}\right]   &
=\epsilon_{i}\nabla\phi^{\ast}(\mathbf{r})\cdot\mathbf{n}\text{ on }%
\partial\Omega_{i},\label{6.9}%
\end{align}
where $\partial$ denotes the boundary of a domain, the jump function
$[\phi(\mathbf{r})]=\lim_{\mathbf{r}_{sh}\rightarrow\mathbf{r}}\phi
(\mathbf{r}_{sh})-\lim_{\mathbf{r}_{i}\rightarrow\mathbf{r}}\phi
(\mathbf{r}_{i})$ at $\mathbf{r}$ $\in\partial\Omega_{i}$ with $\mathbf{r}%
_{sh}\in$ $\Omega_{sh}$ and $\mathbf{r}_{i}\in$ $\Omega_{i}$, $\epsilon
(\mathbf{r})=\epsilon_{s}$ in $\Omega_{sh}$ and $\epsilon(\mathbf{r}%
)=\epsilon_{ion}\epsilon_{0}$ in $\Omega_{i}$, $\epsilon_{ion}$ is a
dielectric constant in $\Omega_{i}$, $\mathbf{n}$ is an outward normal unit
vector at $\mathbf{r}\in$ $\partial\Omega_{i}$, and $\phi^{\ast}%
(\mathbf{r})=q_{i}/(4\pi\epsilon_{i}\left\vert \mathbf{r-0}\right\vert )$. Eq.
(\ref{6.2}) avoids large errors in a direct approximation of the delta
function $\delta(\mathbf{r}-\mathbf{0})$ in the singular charge $q_{i}%
\delta(\mathbf{r}-\mathbf{0})$ of the solvated ion at the origin $\mathbf{0}$
by transforming the singular charge to the Green's function $\phi^{\ast
}(\mathbf{r})$ on $\partial\Omega_{i}$ in (\ref{6.9}) as an approximation
source of the electric field produced by the solvated ion \cite{CL03,GY07}.

For simplicity, we consider a general binary ($K=2$) electrolyte C$_{z_{2}}%
$A$_{z_{1}}$ with the valences of the cation C$^{z_{1}+}$ and anion
A$^{z_{2}-} $ being $z_{1}$ and $z_{2}$, respectively. The first-order Taylor
approximation of the charge density functional $\rho_{ion}(\phi(\mathbf{r}))$
in (\ref{3.7}) with respect to the electric potential $\phi(\mathbf{r})$
yields%
\begin{equation}
\rho_{ion}(\phi(\mathbf{r}))\approx\frac{-C_{1}^{B}q_{1}}{k_{B}T}\left[
\left(  q_{1}-q_{2}\right)  -\Lambda q_{1}\right]  \phi(\mathbf{r}%
),\label{6.10}%
\end{equation}
where $\Lambda=C_{1}^{B}\left(  v_{1}-v_{2}\right)  ^{2}/\left[  \Gamma
^{B}v_{0}+\left(  v_{1}^{2}C_{1}^{B}+v_{2}^{2}C_{2}^{B}+v_{3}^{2}C_{3}%
^{B}\right)  \right]  $ which is a quantity corresponding to a linearization
of the steric potential $S^{trc}(\mathbf{r})$ \cite{LL20}. Consequently, we
obtain a \textbf{generalized Debye length}
\begin{equation}
l_{D4PBik}=\left(  \frac{\epsilon_{s}k_{B}T}{C_{1}^{B}((1-\Lambda)q_{1}%
^{2}-q_{1}q_{2})}\right)  ^{1/2}\label{6.11}%
\end{equation}
that reduces to the original Debye length $l_{D}$ \cite{LM03} if $v_{1}%
=v_{2}\neq0$ (two ionic species having equal radius and thus $\Lambda=0$) or
$v_{1}=v_{2}=$ $v_{3}=0$ (all particles treated as volumeless points in
standard texts for PB \cite{LM03}). The nonlinear value of $\Lambda\neq0 $ for
$v_{1}=v_{2}\neq0$ can be obtained by Newton's method \cite{LL20}.

Eq. (\ref{6.3}) is a second-order PDE that requires two boundary conditions
like (\ref{6.5}) and (\ref{6.6}) for a unique solution $\psi(\mathbf{r})$.
Since $\psi(\mathbf{r})=$ $\epsilon_{s}\nabla^{2}\phi(\mathbf{r}%
)=-\rho(\mathbf{r})\approx\epsilon_{s}\kappa^{2}\phi(\mathbf{r})$ if $l_{c}%
=0$, Eq. (\ref{6.6}) is a simplified (approximate) boundary condition for
$\psi(\mathbf{r})$ on $\partial\Omega_{sh}\cap\partial\Omega_{s}$ without
involving higher-order derivatives of $\psi(\mathbf{r})$ (or the third-order
derivative of $\phi(\mathbf{r})$). The approximations in (\ref{6.6}) and
(\ref{6.10}) do not significantly affect our generalized DH model to fit
activity data. However, these assumptions should be carefully scrutinized in
other applications such as highly charged surfaces. Bazant et al. have
recently developed more consistent and general boundary conditions for their
fourth-order model by enforcing continuity of the Maxwell stress at a charged
interface \cite{DB19,MD19}.

In \cite{LL20}, we analytically solve the linear 4PBik PDEs (\ref{6.2}),
(\ref{6.3}), and (\ref{6.4}) with (\ref{6.10}) in a similar way as Debye and
H\"{u}ckel solved the linear PB equation for a spherically symmetric system.
However, the spherical domain shown in Fig. 1 and the boundary and interface
conditions in (\ref{6.5})-(\ref{6.9}) are different from those of the standard
method for the linear PB equation in physical chemistry texts \cite{LM03}. The
analysis consists of the following steps: (i) The nonlinear term $\rho
_{ion}(\mathbf{r})$ in (\ref{6.3}) is linearized to the linear term
$-\epsilon_{s}\phi/l_{D4PBik}^{2}$ in (\ref{6.10}) as that of Debye and
H\"{u}ckel. (ii) The linear PDEs corresponding to (\ref{6.3}) and (\ref{6.4})
are then formulated into a system of eigenvalue problems with eigenfunctions
$\left(  \phi(\mathbf{r}),\text{ }\psi(\mathbf{r})\right)  $ and eigenvalues
$\left(  \lambda_{1}\text{, }\lambda_{2}\right)  $, where the general solution
of $\phi(\mathbf{r})$ is equal to that of Debye and H\"{u}ckel in the solvent
domain $\Omega_{s}$ (not the entire domain) when $l_{c}=v_{1}=v_{2}=$
$v_{3}=0$. (iii) A unique pair of eigenfunctions $\left(  \phi^{4PBik}%
(\mathbf{r}),\text{ }\psi^{4PBik}(\mathbf{r})\right)  $ is found under
conditions (\ref{6.5})-(\ref{6.9}), where $\phi^{4PBik}(\mathbf{r})$ is equal
to that of Debye and H\"{u}ckel in $\Omega_{s}$ when $l_{c}=v_{1}=v_{2}=$
$v_{3}=0$.

The analytical potential function that we found \cite{LL20} is%
\begin{equation}
\phi^{4PBik}(r)=\left\{
\begin{array}
[c]{l}%
\frac{q_{i}}{4\pi\epsilon_{s}R_{i}^{Born}}+\frac{q_{i}}{4\pi\epsilon_{s}%
R_{i}^{sh}}\left(  \Theta-1\right)  \text{ in }\Omega_{i}\\
\frac{q_{i}}{4\pi\epsilon_{s}r}+\frac{q_{i}}{4\pi\epsilon_{s}R_{i}^{sh}%
}\left(  \Theta-1\right)  \text{ in }\Omega_{sh}\\
\frac{q_{i}}{4\pi\epsilon_{s}r}\left[  \frac{\lambda_{1}^{2}e^{-\sqrt
{\lambda_{2}}\left(  r-R_{i}^{sh}\right)  }-\lambda_{2}^{2}e^{-\sqrt
{\lambda_{1}}\left(  r-R_{i}^{sh}\right)  }}{\lambda_{1}^{2}\left(
\sqrt{\lambda_{2}}R_{i}^{sh}+1\right)  -\lambda_{2}^{2}\left(  \sqrt
{\lambda_{1}}R_{i}^{sh}+1\right)  }\right]  \text{ in }\Omega_{s},
\end{array}
\right. \label{6.12}%
\end{equation}
where%
\begin{equation}
\Theta=\frac{\lambda_{1}^{2}-\lambda_{2}^{2}}{\lambda_{1}^{2}\left(
\sqrt{\lambda_{2}}R_{i}^{sh}+1\right)  -\lambda_{2}^{2}\left(  \sqrt
{\lambda_{1}}R_{i}^{sh}+1\right)  }\text{,}\label{6.13}%
\end{equation}
$r=\left\vert \mathbf{r}\right\vert $, $\lambda_{1}=\left(  1-\sqrt
{1-4l_{c}^{2}/l_{D4PBik}^{2}}\right)  /\left(  2l_{c}^{2}\right)  $, and
$\lambda_{2}=\left(  1+\sqrt{1-4l_{c}^{2}/l_{D4PBik}^{2}}\right)  /\left(
2l_{c}^{2}\right)  $. Note that $\lim_{l_{c}\rightarrow0}\lambda
_{1}=1/l_{D4PBik}^{2}$, $\lim_{l_{c}\rightarrow0}\lambda_{2}=\infty$, and
$\lim_{l_{c}\rightarrow0}\Theta=\lim_{C_{1}^{B}\rightarrow0}\Theta
=\lim_{l_{D4PBik}\rightarrow\infty}\Theta=1$ \cite{LL20}. The linearized 4PBik
potential $\phi^{4PBik}(r)$ reduces to the linearized PB potential $\phi
^{PB}(r)=q_{i}e^{-r/l_{D}}/(4\pi\epsilon_{s}r)$ as in standard texts (e.g. Eq.
(7.46) in \cite{LM03}) by taking $\lim_{l_{c}\rightarrow0}\phi^{4PBik}(r)$
with $v_{k}=0$ for all $k$, $R_{i}^{sh}=0$, and $r>0$ \cite{LL20}.

As discussed in \cite{VB15}, since the solvation free energy of an ion $i$
varies with salt concentrations, the Born energy $q_{i}^{2}\left(  \frac
{1}{\epsilon_{w}}-1\right)  /8\pi\epsilon_{0}R_{i}^{0}$ in pure water (i.e.
$C_{i}^{B}=0$) with a constant Born radius $R_{i}^{0}$ should change to depend
on $C_{i}^{B}\geq0$. Equivalently, the Born radius $R_{i}^{Born}$ in
(\ref{6.12}) is variable and we can model it from $R_{i}^{0}$ by a simple
formula \cite{LE15a,LE18}
\begin{equation}
R_{i}^{Born}=\theta R_{i}^{0}\text{, \ \ }\theta=1+\alpha_{1}^{i}\left(
\overline{C}_{i}^{B}\right)  ^{1/2}+\alpha_{2}^{i}\overline{C}_{i}^{B}%
+\alpha_{3}^{i}\left(  \overline{C}_{i}^{B}\right)  ^{3/2},\label{6.14}%
\end{equation}
where $\overline{C}_{i}^{B}=$ $C_{i}^{B}$/M is a dimensionless bulk
concentration and $\alpha_{1}^{i}$, $\alpha_{2}^{i}$, and $\alpha_{3}^{i}$ are
parameters for modifying the experimental Born radius $R_{i}^{0}$ to fit
experimental activity coefficient $\gamma_{i}$ that changes with the bulk
concentration $C_{i}^{B}$ of the ion. The Born radii $R_{i}^{0}$ given below
are from \cite{VB15} obtained from the experimental hydration Helmholtz free
energies of those ions given in \cite{F04}. The three parameters in
(\ref{6.14}) have physical or mathematical meanings unlike many parameters in
the Pitzer model \cite{F10,V11,RK15}. The first parameter $\alpha_{1}^{i}$
adjusts $R_{i}^{0}$ and accounts for the real thickness of the ionic
atmosphere (Debye length), which is proportional to the square root of the
ionic strength in the DH theory \cite{LM03}. The second $\alpha_{2}^{i}$ and
third $\alpha_{3}^{i}$ parameters are adjustments in the next orders of
approximation beyond the DH treatment of ionic atmosphere \cite{LE18}.

The potential value $\phi^{0}(\mathbf{0})=\lim_{C_{1}^{B}\rightarrow0}%
\phi^{4PBik}(\mathbf{0})=$ $q_{i}/\left(  4\pi\epsilon_{s}R_{i}^{0}\right)  $
by $\lim_{C_{1}^{B}\rightarrow0}\Theta=1$ and $\lim_{C_{1}^{B}\rightarrow
0}R_{i}^{Born}=R_{i}^{0}$. From (\ref{6.1}) and (\ref{6.12}), we thus have a
\textbf{generalized activity coefficient} $\gamma_{i}^{4PBik}$ in%
\begin{equation}
\ln\gamma_{i}^{4PBik}=\frac{q_{i}^{2}}{8\pi\epsilon_{s}k_{B}T}\left(  \frac
{1}{R_{i}^{Born}}-\frac{1}{R_{i}^{0}}+\frac{\Theta-1}{R_{i}^{sh}}\right)
,\label{6.15}%
\end{equation}
which satisfies the DH limiting law, i.e., $\gamma_{i}^{4PBik}=\gamma_{i}%
^{DH}=1$ for infinite dilute (ideal) solutions as $C_{i}^{B}\rightarrow0$. The
generalized activity coefficient $\gamma_{i}^{4PBik}$ reduces to the classical
DH activity coefficient $\gamma_{i}^{DH}$ \cite{DH23}, namely,
\begin{equation}
\ln\gamma_{i}^{DH}=\frac{-q_{i}^{2}}{8\pi\epsilon_{s}k_{B}T(R_{i}+l_{D}%
)}\label{6.16}%
\end{equation}
if $R_{i}^{Born}=R_{i}^{0}$ (without considering Born energy effects),
$R_{i}^{sh}=R_{i}$ (an effective ionic radius \cite{DH23}), $l_{D4PBik}=l_{D}$
(no steric effect), and $l_{c}=0$ (no correlation effect). The reduction shown
in \cite{LL20} is by taking the limit of the last term in (\ref{6.15}) as
$l_{c}\rightarrow0$, i.e., $\lim_{l_{c}\rightarrow0}\frac{\Theta-1}{R_{i}%
^{sh}}=\frac{-1}{R_{i}+l_{D}}$.

\section{Numerical Methods}

Numerical simulations are indispensable to study chemical, physical, and
mathematical properties of biological and chemical systems in realistic
applications, especially with experimental details at atomic scale such as ion
channels in the Protein Data Bank (PDB) \cite{B00}. Continuum PDE models have
substantial advantages over Monte Carlo, Brownian dynamics (BD), or molecular
dynamics in physical insights and computational efficiency that are of great
importance in studying a range of conditions and concentrations especially for
large nonequilibrium or inhomogeneous systems, as are present in experiments
and in life itself
\cite{FB02,E11,WZ12,SK10,GK04,LH10,ZC11,IR02,E10,E12,E13,BF14,KM15,LG17,CF19}.

The literature on numerical methods for solving PB and PNP models is vast
\cite{L13,LE15,CC18}. We summarize here some important features of the methods
proposed in \cite{L13,LE15,CC18} for Poisson-Bikerman and
Poisson-Nernst-Planck-Bikerman models, which may be useful for workers in
numerical analysis and coding practice. Since PNPB including 4PBik is highly
nonlinear and the geometry of protein structures is very complex, we emphasize
two different types of methods, namely, nonlinear iterative methods and
discretization methods for these two problems as follows.

\subsection{Nonlinear Iterative Methods}

For the PNPB system of $K+1$ NP Eqs. (\ref{5.1}), Laplace Eq. (\ref{6.2}), and
two 4PBik Eqs. (\ref{6.3}) and (\ref{6.4}), the total number of second-order
PDEs that we need to solve is $K+4$. These PDEs are coupled together and
highly nonlinear except (\ref{6.2}). Numerically solving this kind of
nonlinear systems is not straightforward \cite{L13,LE15,CC18}. We use the
following algorithm to explain essential procedures for solving the
steady-state PNPB system, where $\Omega_{m}$ denotes the biomolecular domain
that contains a total of $Q$ fixed atomic charges $q_{j}$ located at
$\mathbf{r}_{j}$ in a channel protein as shown in Fig. 2 for the gramicidin A
channel downloaded from PDB with $Q=554$, for example, $\partial\Omega_{m}$
denotes the molecular surface of the protein and the membrane lipids through
which the protein crosses as shown in Fig. 3, and $\Omega_{s}$ is the solvent
domain consisting of the channel pore and the extracellular and intracellular
baths for mobile ions and water molecules. \begin{figure}[t]
\centering\includegraphics[scale=1.0]{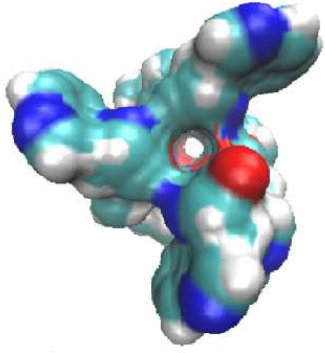}\caption{Top view of the
gramicidin A channel.}%
\end{figure}\begin{figure}[tt]
\centering\includegraphics[scale=1.0]{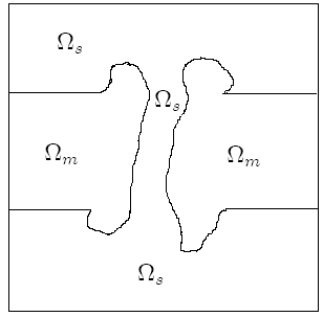}\caption{A cross section of 3D
simulation domain for the channel placed in a rectangular box, where
$\Omega_{m}$ is the biomolecular domain consisting of the channel protein and
the membrane and $\Omega_{s}$ is the solvent domain consisting of the channel
pore and the baths.}%
\end{figure}

\textit{Nonlinear Iterative Algorithm \cite{LE15}:}

\begin{enumerate}
\item Solve Laplace Eq. $-\nabla^{2}\phi(\mathbf{r})=0$ for $\phi
^{0}(\mathbf{r})$ in $\Omega_{m}$ once for all with $\phi^{0}(\mathbf{r}%
)=\phi^{\ast}(\mathbf{r})=\sum_{j=1}^{Q}q_{j}/(4\pi\epsilon_{m}\epsilon
_{0}\left\vert \mathbf{r-r}_{j}\right\vert )$ on $\partial\Omega_{m}$.

\item Solve Poisson Eq. $-\nabla\cdot\left(  \epsilon\nabla\phi(\mathbf{r}%
)\right)  =\rho_{ion}(\mathbf{r})$ for $\phi^{Old}(\mathbf{r})$ in $\Omega
_{s}$ with $\rho_{ion}(\mathbf{r})=0$, $\phi^{Old}=V=0$ on $\partial\Omega$,
and the jump condition $\left[  \epsilon\nabla\phi^{Old}\cdot\mathbf{n}%
\right]  =-\epsilon_{m}\epsilon_{0}\nabla(\phi^{\ast}+\phi^{0})\cdot
\mathbf{n}$ on $\partial\Omega_{m}$ as (\ref{6.9}), where $V$ denotes applied voltage.

\item $V=V_{0}\neq0$ an initial voltage.

\item Solve PF1 Eq. $\epsilon_{s}\left(  \lambda_{c}l_{c}^{2}\nabla
^{2}-1\right)  \Psi(\mathbf{r})=\sum_{i=1}^{K}q_{i}C_{i}^{Old}(\mathbf{r})$
for $\Psi^{New}(\mathbf{r})$ in $\Omega_{s}$ with $\nabla\Psi^{New}%
\cdot\mathbf{n}=0$ on $\partial\Omega_{m}$, $\Psi^{New}=0$ on $\partial\Omega
$, $C_{i}^{Old}(\mathbf{r})=C_{i}^{B}\exp\left(  -\beta_{i}\phi^{Old}%
(\mathbf{r})+\frac{v_{i}}{v_{0}}S^{trc}(\mathbf{r})\right)  $,\ $S^{trc}%
(\mathbf{r})=\ln\frac{\Gamma^{Old}(\mathbf{r)}}{\Gamma^{B}}$, and
$\Gamma^{Old}(\mathbf{r)}=1-\sum_{j=1}^{K+1}\lambda_{s}v_{j}C_{j}%
^{Old}(\mathbf{r}) $.

\item Solve PF2 Eq. $-\nabla\cdot\left(  \epsilon_{s}\nabla\phi(\mathbf{r}%
)\right)  +\rho^{\prime}(\phi^{Old})\phi(\mathbf{r})=-\epsilon\Psi^{New}%
+\rho^{\prime}(\phi^{Old})\phi^{Old}$ for $\phi^{New}(\mathbf{r})$ in
$\Omega_{s}$ with $\phi^{New}=V$ on $\partial\Omega$ and the same jump
condition in Step 2, where $\rho^{\prime}(\phi)$ is the derivative of
$\rho(\phi)$ with respect to $\phi$.

\item If the maximum error norm $\left\Vert \phi^{New}-\phi^{Old}\right\Vert
_{\infty}>Tol$, a preset tolerance, then set $\phi^{Old}=\omega_{4PBik}%
\phi^{Old}+(1-\omega_{4PBik})\phi^{New}$ and go to Step 4, else go to Step 7.

\item Solve NP Eq. $-\nabla\cdot\mathbf{J}_{i}(\mathbf{r})=0$ for $C_{i}%
^{New}(\mathbf{r})$ in $\Omega_{s}$ for all $i=1,\cdots,K+1$ with
$\mathbf{J}_{i}(\mathbf{r})=-D_{i}\left[  \nabla C_{i}(\mathbf{r})+\beta
_{i}C_{i}(\mathbf{r})\right.  $ $\left.  \nabla\phi^{Old}(\mathbf{r}%
)-\lambda_{s}\frac{v_{i}}{v_{0}}C_{i}(\mathbf{r})\nabla S^{trc}(\mathbf{r}%
)\right]  $,\ $S^{trc}(\mathbf{r})=\ln\frac{\Gamma^{Old}(\mathbf{r)}}%
{\Gamma^{B}}$, $C_{i}^{New}(\mathbf{r})=0$ on $\partial\Omega$, and
$\mathbf{J}_{i}(\mathbf{r})\cdot\mathbf{n}=0$ on $\partial\Omega_{m}$.

\item Solve PF1 Eq. for $\Psi^{New}$ as in Step 4 with $C_{i}^{New}$ in place
of $C_{i}^{Old}$.

\item Solve PF2 Eq. for $\phi^{New}$ as in Step 5.

\item If $\left\Vert \phi^{New}-\phi^{Old}\right\Vert _{\infty}>Tol$, then set
$\phi^{Old}=\omega_{PNPB}\phi^{Old}+(1-\omega_{PNPB})\phi^{New} $ and go to
Step 7, else go to Step 11.

\item $V=V+\Delta V$ and go to Step 4 until the desired voltage is reached.
\end{enumerate}

Linearizing the nonlinear 4PBik (\ref{3.7}) yields two second-order linear PF1
and PF2 in Steps 4 and 5 that differ from the nonlinear (\ref{6.3}) and
(\ref{6.4}). Newton's iterative Steps 4-6 for solving PF1 and PF2 dictates
convergence that also depends on various mappings from an old solution
$\phi^{Old}$ to a new solution $\phi^{New}$. This algorithm uses two
relaxation and three continuation mappings for which we need to carefully tune
two relaxation parameters $\omega_{4PBik}$ and $\omega_{PNPB}$ and three
continuation parameters $\lambda_{c}$ (related to correlation effects),
$\lambda_{s}$ (steric effects), and $\Delta V$ (incremental voltage for
applied voltage). For example, the parameter $\lambda_{s}$ in $\Gamma
^{Old}(\mathbf{r)}=1-\sum_{j=1}^{K+1}\lambda_{s}v_{j}C_{j}^{Old}(\mathbf{r}) $
can be chosen as $\lambda_{s}=k\Delta\lambda$, $k=0,1,2,\cdots,\frac{1}%
{\Delta\lambda}$, an incremental continuation from 0 (no steric effects) to 1
(fully steric effects) with a tuning stepping length $\Delta\lambda$. The
algorithm can fail to converge if we choose $\Delta\lambda=1$ (without
continuation) for some simulation cases, since we may have $\Gamma
^{Old}(\mathbf{r})<0$ resulting in numerically \textbf{undefined}
$S^{trc}(\mathbf{r})=\ln\frac{\Gamma^{Old}(\mathbf{r)}}{\Gamma^{B}}$ at some
$\mathbf{r}$ where the potential $\phi^{Old}(\mathbf{r})$ is large.

\subsection{Discretization Methods}

All PDEs in Steps 1, 2, 4, 5, 8, and 9 are of Poisson type $-\nabla^{2}%
\phi(\mathbf{r})=f(\mathbf{r})$. We use the central finite difference (FD)
method \cite{L13}%
\begin{equation}
\left.
\begin{array}
[c]{l}%
\frac{-\phi_{i-1,j,k}+2\phi_{ijk}-\phi_{i+1,j,k}}{\Delta x^{2}}+\frac
{-\phi_{i,j-1,k}+2\phi_{ijk}-\phi_{i,j+1,k}}{\Delta y^{2}}+\\
\frac{-\phi_{i,j,k-1}+2\phi_{ijk}-\phi_{i,j,k+1}}{\Delta z^{2}}=f_{ijk},
\end{array}
\right.  \label{7.1}%
\end{equation}
to discretize it at all grid points $\mathbf{r}_{ijk}=(x_{i},y_{j},z_{k})$ in
a domain, where $\phi_{ijk}\approx\phi(x_{i},y_{j},z_{k})$, $f_{ijk}%
=f(x_{i},y_{j},z_{k})$, and $\Delta x$, $\Delta y$, and $\Delta z$ are mesh
sizes on the three axes from a uniform partition $\Delta x=\Delta y=\Delta
z=h$. The domains in Steps 1 and 2 are $\Omega_{m}$ and $\Omega_{s}$,
respectively. The discretization leads to a sparse matrix system
$A\overrightarrow{\phi}=\overrightarrow{f}$ with the compressed bandwidth of
the matrix $A$ being 7, where the matrix size can be millions for sufficiently
small $h$ to obtain sufficiently accurate $\phi_{ijk}$.

The matrix system consists of four subsystems, two by the FD method
(\ref{7.1}) in $\Omega_{m}$ and $\Omega_{s}$, one by another method (see
below) to discretize the jump condition in Step 2 on the interface
$\partial\Omega_{m}$ between $\Omega_{s}$ and $\Omega_{m}$, and one by
imposing boundary conditions on $\partial\Omega$. We need to solve the matrix
system in the entire domain $\overline{\Omega}=\overline{\Omega}_{m}%
\cup\overline{\Omega}_{s}$.

The convergence order of (\ref{7.1}) is $O(h^{2})$ (optimal) in maximum error
norm for sufficiently smooth function $\phi(\mathbf{r})$. However, this
optimal order can be easily degraded to $O(h^{0.37})$ \cite{HL05}, for
example, by geometric singularities if the jump condition is not properly
treated. In \cite{L13}, we propose the interface method%
\begin{equation}
\frac{-\epsilon_{i-\frac{3}{2}}\phi_{i-2}+\left(  \epsilon_{i-\frac{3}{2}%
}+\left(  1-A_{1}\right)  \epsilon_{i-\frac{1}{2}}^{-}\right)  \phi
_{i-1}-A_{2}\epsilon_{i-\frac{1}{2}}^{-}\phi_{i}}{\Delta x^{2}}=f_{i-1}%
+\frac{\epsilon_{i-\frac{1}{2}}^{-}A_{0}}{\Delta x^{2}} \label{7.2}%
\end{equation}%
\begin{equation}
\frac{-B_{1}\epsilon_{i-\frac{1}{2}}^{+}\phi_{i-1}+\left(  \left(
1-B_{2}\right)  \epsilon_{i-\frac{1}{2}}^{+}+\epsilon_{i+\frac{1}{2}}\right)
\phi_{i}-\epsilon_{i+\frac{1}{2}}\phi_{i+1}}{\Delta x^{2}}=f_{i}%
+\frac{\epsilon_{i-\frac{1}{2}}^{+}B_{0}}{\Delta x^{2}}, \label{7.3}%
\end{equation}
where%
\[
A_{1}=\frac{-\left(  \epsilon_{m}-\epsilon_{s}\right)  }{\epsilon_{m}%
+\epsilon_{s}},\text{ }A_{2}=\frac{2\epsilon_{m}}{\epsilon_{m}+\epsilon_{s}%
},\text{ }A_{0}=\frac{-2\epsilon_{m}\left[  \phi\right]  -\Delta x\left[
\epsilon\phi^{\prime}\right]  }{\epsilon_{m}+\epsilon_{s}},
\]%
\[
B_{1}=\frac{2\epsilon_{s}}{\epsilon_{m}+\epsilon_{s}},\text{ }B_{2}%
=\frac{\epsilon_{m}-\epsilon_{s}}{\epsilon_{m}+\epsilon_{s}},\text{ }%
B_{0}=\frac{2\epsilon_{s}\left[  \phi\right]  -\Delta x\left[  \epsilon
\phi^{\prime}\right]  }{\epsilon_{m}+\epsilon_{s}},
\]
to discretize the 1D Poisson equation $-\frac{d}{dx}\left(  \epsilon
(x)\frac{d\phi(x)}{dx}\right)  =f(x)$ at every jump position $\gamma
\in\partial\Omega_{m}$ that is at the middle of its two neighboring grid
points, i.e., $x_{i-1}<\gamma=x_{i-\frac{1}{2}}<x_{i}$, where $x_{i-\frac
{1}{2}}=(x_{i-1}+x_{i})/2$ and $x_{i-1}$ and $x_{i}$ belong to different
domains $\Omega_{s}$ and $\Omega_{m}$. The corresponding cases in $y$- and
$z$-axis follow obviously in a similar way. This method yields
\textbf{optimal} convergence \cite{L13}.

Since the matrix system is usually very large in 3D simulations and we need to
repeatedly solve such systems updated by nonlinear iterations as shown in the
above algorithm, linear iterative methods such as the bi-conjugate gradient
stabilized (bi-CG) method are used to solve the matrix system \cite{CC18}. We
propose two parallel algorithms (one for bi-CG and the other for nonlinear
iterations) in \cite{CC18} and show that parallel algorithms on GPU (graphic
processing unit) over sequential algorithms on CPU (central processing unit)
can achieve 22.8$\times$ and 16.9$\times$ speedups for the linear solver time
and total runtime, respectively.

Discretization of Nernst-Planck Eq. in Step 7 is different from (\ref{7.1})
because the standard FD method%
\begin{equation}
\frac{C_{i+1}-C_{i}}{\Delta x}=\frac{C_{i+1}+C_{i}}{2}\left(  -\beta
\frac{\Delta\phi_{i}}{\Delta x}+\frac{\Delta S_{i}^{trc}}{\Delta x}\right)
\label{7.4}%
\end{equation}
for the zero flux ($J(x)=-D(x)\left(  \frac{dC(x)}{dx}+\beta C(x)\frac
{d\phi(x)}{dx}-\frac{v}{v_{0}}C(x)\frac{dS^{trc}(x)}{dx}\right)  =0$) can
easily yield%
\begin{equation}
C_{i+1}-C_{i}>C_{i+1}+C_{i} \label{7.5}%
\end{equation}
and thereby a negative (\textbf{unphysical}) concentration $C_{i}<0$ at
$x_{i}$ if%
\begin{equation}
\frac{1}{2}\left(  -\beta\Delta\phi_{i}+\Delta S_{i}^{trc}\right)  >1,
\label{7.6}%
\end{equation}
where $\Delta\phi_{i-1}=\phi_{i}-\phi_{i-1}$, $\phi_{i}\approx\phi(x_{i})$
etc. Therefore, it is crucial to check whether the \textbf{generalized
Scharfetter-Gummel} (SG) condition \cite{LE15}%
\begin{equation}
-\beta\Delta\phi_{i}+\Delta S_{i}^{trc}\leq2 \label{7.7}%
\end{equation}
is satisfied by any discretization method in implementation. This condition
generalizes the the well-known SG stability condition in semiconductor device
simulations \cite{SG69,S88} to include the steric potential function
$S^{trc}(\mathbf{r})$.

We extend the classical SG method \cite{SG69} of the flux $J(x)$ in
\cite{LE15} to
\begin{equation}
J_{i+\frac{1}{2}}=\frac{-D}{\Delta x}\left[  B(-t_{i})C_{i+1}-B(t_{i}%
)C_{i}\right]  \label{7.8}%
\end{equation}
where $t_{i}=\beta\Delta\phi_{i}-\Delta S_{i}^{trc}$ and $B(t)=\frac{t}%
{e^{t}-1}$ is the Bernoulli function \cite{S88}. Eq. (\ref{7.8}), an
exponential fitting scheme, satisfies (\ref{7.7}) and is derived from assuming
that the flux $J$, the local electric field $\frac{-d\phi}{dx}$, and the local
steric field $\frac{dS^{trc}}{dx}$ are all constant in the sufficiently small
subinterval $(x_{i}$, $x_{i+1})$, i.e.,
\begin{equation}
\frac{J}{D}=\frac{-dC(x)}{dx}-kC(x)\text{, for all }x\in(x_{i}\text{, }%
x_{i+1})\text{,}\label{7.9}%
\end{equation}
where $k=\beta\frac{d\phi}{dx}-\frac{dS^{trc}}{dx}$. Solving this ordinary
differential equation (ODE) with a boundary condition $C_{i}$ or $C_{i+1}$
yields the well-known Goldman-Hodgkin-Katz flux equation in ion channels
\cite{H01}, which is exactly the same as that in (\ref{7.8}) but with the
subinterval $(x_{i}$, $x_{i+1})$ being replaced by the height of the entire
box in Fig. 3.

The generalized Scharfetter-Gummel method for Nernst-Planck Eq. is thus%
\begin{align}
\frac{dJ(x_{i})}{dx}  &  \approx\frac{J_{i+\frac{1}{2}}-J_{i-\frac{1}{2}}%
}{\Delta x}=\frac{a_{i-1}C_{i-1}+a_{i}C_{i}+a_{i+1}C_{i+1}}{\Delta x^{2}%
}=0\label{7.10}\\
J_{i-\frac{1}{2}}  &  =\frac{-D}{\Delta x}\left[  B(-t_{i-1})C_{i}%
-B(t_{i-1})C_{i-1}\right] \nonumber\\
J_{i+\frac{1}{2}}  &  =\frac{-D}{\Delta x}\left[  B(-t_{i})C_{i+1}%
-B(t_{i})C_{i}\right] \nonumber\\
t_{i}  &  =\beta\Delta\phi_{i}-\Delta S_{i}^{trc}\text{, }B(t)=\frac{t}%
{e^{t}-1}\nonumber\\
a_{i-1}  &  =-B(t_{i-1})\text{, }a_{i}=B(-t_{i-1})+B(t_{i})\text{, }%
a_{i+1}=-B(-t_{i})\text{. }\nonumber
\end{align}
The SG method is \textbf{optimal} in the sense that it integrates the ODE
(\ref{7.9}) \textit{exactly} at \textit{every} grid point with a suitable
boundary condition \cite{MR83}. Therefore, the SG method can resolve sharp
layers very accurately \cite{MR83} and hence needs few grid points to obtain
tolerable approximations when compared with the primitive FD method. Moreover,
the exact solution of (\ref{7.9}) for the concentration function $C(x)$ yields
an exact flux $J(x)$. Consequently, the SG method is \textbf{current
preserving}, which is particularly important in nonequilibrium systems, where
the current is possibly the most relevant physical property of interest
\cite{BM89}.

It is difficult to overstate the importance of the current preserving feature
and it must be emphasized for workers coming from fluid mechanics that
preserving current has a significance quite beyond the preserving of flux in
uncharged systems. Indeed, conservation of current (defined as Maxwell did to
include the displacement current of the vacuum $\epsilon_{0}\frac
{\partial\mathbf{E}(\mathbf{r},t)}{\partial t}$) is an unavoidable
consequence, nearly a restatement of the Maxwell equations themselves
\cite{EO17,E19}. The electric field is so strong that the tiniest error in
preserving current, i.e., the tiniest deviation from Maxwell's equations,
produces huge effects. The third paragraph of Feynman's lectures on
electrodynamics makes this point unforgettable \cite{FL63}. Thus, the
consequences of a seemingly small error in preserving the flow of charge are
dramatically larger than the consequences of the same error in preserving the
flux of mass.

\section{Results}

We have used the saturating Poisson-Nernst-Plack-Bikerman theory to study ion
activities, electric double layers, and biological ion channels in the past.
The theory accounts for the steric effect of ions and water molecules, the
effects of ion-ion and ion-water correlations, the screening and polarization
effects of polar water, and the charge/space competition effect of ions and
water molecules of different sizes and valences. These effects are all closely
related to the dielectric operator in (\ref{3.7}) and the steric potential in
(\ref{2.4}) that works on both macroscopic and atomic scales. We now
illustrate these properties in the following three areas using mostly
experimental data to verify the theory.

\subsection{Ion Activities}

The curves in Fig. 4 obtained by the generalized Debye-H\"{u}ckel formula
(\ref{6.15}) \cite{LL19} fit well to the experimental data by Wilczek-Vera et
al. \cite{WR04} for single-ion activities in 8 1:1 electrolytes. There are
only three fitting parameters in the formula, namely $\alpha_{1}^{i}$,
$\alpha_{2}^{i}$ and $\alpha_{3}^{i}$, which we reiterate have specific
physical meaning as parameters of the water shell around ions. The values of
the parameters are given in Table 1 from which we observe that $R_{i}^{Born}$
deviates from $R_{i}^{0}$ slightly. For example, $R_{_{\text{Cl}^{-}}}%
^{Born}/R_{\text{Cl}^{-}}^{0}=1.007\sim1.044$ (not shown) for Fig. 4a with
[LiCl] = $0\sim2.5$ M, since the cavity radius $R_{_{\text{Cl}^{-}}}^{Born}$
is an atomic measure from the infinite singularity $\delta(\mathbf{r}%
-\mathbf{0})$ at the origin, i.e., $\phi^{4PBik}(r)$ and thus $\gamma
_{i}^{4PBik}$ are very sensitive to $R_{i}^{Born}$. On the other hand,
$\gamma_{i}^{4PBik}$ is not very sensitive to $R_{i}^{sh}$ ($R_{\text{Cl}^{-}%
}^{sh}=5.123\sim5.083$ \r{A}), i.e., the fixed choice of $O_{i}^{\text{w}}=18$
(an experimental value in \cite{RI13}) for all curves is not critical but
reasonable \cite{LE15a}. The error between the estimated $O_{i}^{\text{w}}$
and its unknown true value can always be compensated by small adjustments of
$R_{i}^{Born}$. Table 1 also shows the significant order of these parameters,
i.e., $\left\vert \alpha_{1}^{i}\right\vert >\left\vert \alpha_{2}%
^{i}\right\vert >\left\vert \alpha_{3}^{i}\right\vert $ in general cases. The
values of other symbols are $a_{\text{Li}^{+}}=0.6$, $a_{\text{Na}^{+}}=0.95$,
$a_{\text{K}^{+}}=1.33$, $a_{\text{F}^{-}}=1.36$, $a_{\text{Cl}^{-}}=1.81$,
$a_{\text{Br}^{-}}=1.95$, $a_{\text{H}_{2}\text{O}}=1.4$ \r{A}, $R_{\text{Li}%
^{+}}^{0}=1.3$, $R_{\text{Na}^{+}}^{0}=1.618$, $R_{\text{K}^{+}}^{0}=1.95$,
$R_{\text{F}^{-}}^{0}=1.6$, $R_{\text{Cl}^{-}}^{0}=2.266$, $R_{\text{Br}^{-}%
}^{0}=2.47$ \r{A}, $\epsilon_{w}=78.45$, $\epsilon_{ion}=1$, $T=298.15$ K.
\begin{figure}[t]
\centering\includegraphics[scale=0.6]{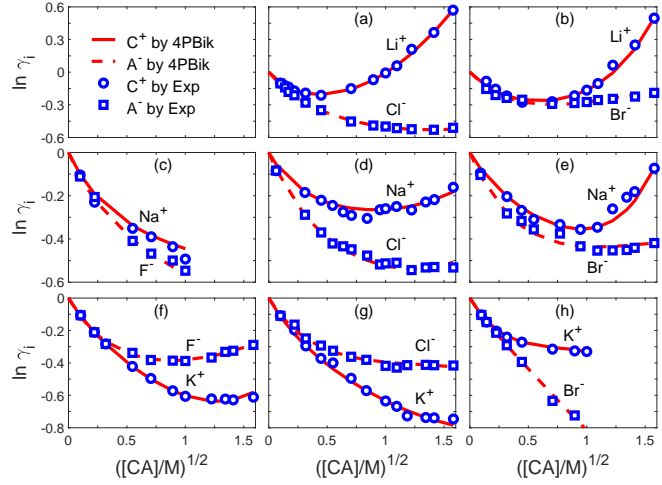}\caption{Single-ion activity
coefficients of 1:1 electrolytes. Comparison of 4PBik results (curves) with
experimental data (symbols) \cite{WR04} on $i=$ C$^{+}$ (cation) and A$^{-}$
(anion) activity coefficients $\gamma_{i}$ in various [CA] from 0 to 1.6 M.}%
\end{figure}

\begin{center}
$%
\begin{tabular}
[c]{ccccc|lcccc}%
\multicolumn{10}{c}{Table 1. Values of $\alpha_{1}^{i}$, $\alpha_{2}^{i}$,
$\alpha_{3}^{i}$ in (\ref{6.14})}\\\hline
Fig.\# & $i$ & $\alpha_{1}^{i}$ & $\alpha_{2}^{i}$ & \ $\alpha_{3}^{i}$ &
Fig.\# & $i$ & $\alpha_{1}^{i}$ & $\alpha_{2}^{i}$ & \ $\alpha_{3}^{i}%
$\\\hline
\multicolumn{1}{l}{3a} & \multicolumn{1}{l}{Li$^{+}$} &
\multicolumn{1}{r}{$-0.006$} & \multicolumn{1}{r}{$-0.037$} &
\multicolumn{1}{r|}{$0.004$} & 3e & \multicolumn{1}{l}{Na$^{+}$} &
\multicolumn{1}{r}{$-0.049$} & \multicolumn{1}{r}{$0.042$} &
\multicolumn{1}{r}{$-0.013$}\\
\multicolumn{1}{l}{3a} & \multicolumn{1}{l}{Cl$^{-}$} &
\multicolumn{1}{r}{$0.052$} & \multicolumn{1}{r}{$-0.015$} &
\multicolumn{1}{r|}{$0$} & 3e & \multicolumn{1}{l}{Br$^{-}$} &
\multicolumn{1}{r}{$0.071$} & \multicolumn{1}{r}{$-0.048$} &
\multicolumn{1}{r}{$0.006$}\\
\multicolumn{1}{l}{3b} & \multicolumn{1}{l}{Li$^{+}$} &
\multicolumn{1}{r}{$-0.006$} & \multicolumn{1}{r}{$-0.011$} &
\multicolumn{1}{r|}{$-0.004$} & 3f & \multicolumn{1}{l}{K$^{+}$} &
\multicolumn{1}{r}{$0.005$} & \multicolumn{1}{r}{$0.051$} &
\multicolumn{1}{r}{$-0.015$}\\
\multicolumn{1}{l}{3b} & \multicolumn{1}{l}{Br$^{-}$} &
\multicolumn{1}{r}{$0.026$} & \multicolumn{1}{r}{$-0.057$} &
\multicolumn{1}{r|}{$0.010$} & 3f & \multicolumn{1}{l}{F$^{-}$} &
\multicolumn{1}{r}{$0.033$} & \multicolumn{1}{r}{$-0.028$} &
\multicolumn{1}{r}{$0.003$}\\
\multicolumn{1}{l}{3c} & \multicolumn{1}{l}{Na$^{+}$} & \multicolumn{1}{r}{$0
$} & \multicolumn{1}{r}{$0$} & \multicolumn{1}{r|}{$0$} & 3g &
\multicolumn{1}{l}{K$^{+}$} & \multicolumn{1}{r}{$0.031$} &
\multicolumn{1}{r}{$0.022$} & \multicolumn{1}{r}{$-0.005$}\\
\multicolumn{1}{l}{3c} & \multicolumn{1}{l}{F$^{-}$} &
\multicolumn{1}{r}{$0.027$} & \multicolumn{1}{r}{$0$} &
\multicolumn{1}{r|}{$0$} & 3g & \multicolumn{1}{l}{Cl$^{-}$} &
\multicolumn{1}{r}{$0.020$} & \multicolumn{1}{r}{$-0.025$} &
\multicolumn{1}{r}{$0.004$}\\
\multicolumn{1}{l}{3d} & \multicolumn{1}{l}{Na$^{+}$} &
\multicolumn{1}{r}{$-0.045$} & \multicolumn{1}{r}{$0.009$} &
\multicolumn{1}{r|}{$-0.002$} & 3h & \multicolumn{1}{l}{K$^{+}$} &
\multicolumn{1}{r}{$0.025$} & \multicolumn{1}{r}{$-0.062$} &
\multicolumn{1}{r}{$0.018$}\\
\multicolumn{1}{l}{3d} & \multicolumn{1}{l}{Cl$^{-}$} &
\multicolumn{1}{r}{$0.063$} & \multicolumn{1}{r}{$-0.014$} &
\multicolumn{1}{r|}{$-0.002$} & 3h & \multicolumn{1}{l}{Br$^{-}$} &
\multicolumn{1}{r}{$0.001$} & \multicolumn{1}{r}{$0.082$} &
\multicolumn{1}{r}{$0$}\\\hline
\end{tabular}
$
\end{center}

The electric potential and other physical properties of ionic activity can be
studied in detail according to the partitioned domain in Fig. 1 characterized
by $R_{i}^{Born}$ and $R_{i}^{sh}$. For example, we observe from Fig. 5 that
the electric potential ($\phi_{\text{Br}^{-}}^{4PBik}(0)=-2.4744$ $k_{B}T/e$)
and the Born radius ($R_{\text{Br}^{-}}^{Born}($2 M$)=2.0637$ \r{A}) generated
by Br$^{-}$ at [LiBr] = 2 M are significantly different from that
($\phi_{\text{Br}^{-}}^{4PBik}(0)=-0.6860$ $k_{B}T/e$, $R_{\text{Br}^{-}%
}^{Born}($2 M$)=4.2578$ \r{A}) at [KBr] = 2 M. The only difference between
these two solutions is the size of cations, i.e., the size of different
positive ions changes significantly the activity of the same negative ion at
high concentrations. The difference between $\phi_{\text{Li}^{+}}^{4PBik}(0)$
and $\phi_{\text{K}^{+}}^{4PBik}(0)$ is due to the sizes of Li$^{+}$ and
K$^{+}$ not Br$^{-}$ as it is the same for both solutions. \begin{figure}[t]
\centering\includegraphics[scale=0.7]{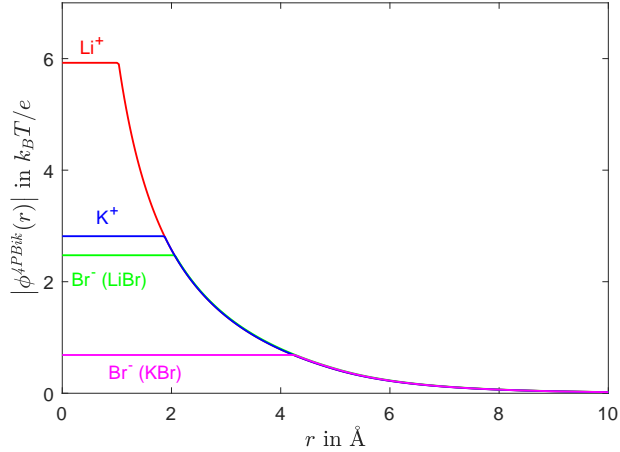}\caption{Electric potential
$\phi^{4PBik}(r)$ profiles by ( \ref{6.12}) near the solvated ions Li$^{+}$
and Br$^{-}$ at [LiBr] = 2, and K$^{+}$ and Br$^{-}$ at [KBr] = 2 M, where $r$
is the distance from the center of the respective ion.}%
\end{figure}

This example clearly shows the atomic properties of 4PBik theory in the ion
$\Omega_{i}$ and shell $\Omega_{sh}$ domains and the continuum properties in
the solvent domain $\Omega_{s}$. The Born radius $R_{i}^{Born}$ in
(\ref{6.12}) determined by (\ref{6.14}) changes with (i) \textbf{ion-water}
interactions in $\Omega_{i}\cup\Omega_{sh}$ and (ii) \textbf{ion-ion}
interactions in $\Omega_{i}\cup\Omega_{s}$ via $\phi^{4PBik}(r)$ in
(\ref{6.12}) that is self-consistently determined by the interface conditions
in (\ref{6.5})-(\ref{6.9}) and by (iii) \textbf{multi-salt }\cite{LE18,LL20}%
\textbf{\ }concentrations in $\Omega_{s}$, (iv) the \textbf{screening} effects
of water in $\Omega_{sh}$ and ions and water in $\Omega_{s}$, (v) the
\textbf{polarization} effect of water in $\Omega_{s}$, (vi) the
\textbf{correlation} effect between ions in $\Omega_{s}$, (vii) the
\textbf{steric} effects of all ions and water in the entire domain
$\overline{\Omega}=\overline{\Omega}_{i}\cup\overline{\Omega}_{sh}%
\cup\overline{\Omega}_{s} $, (viii) \textbf{temperatures} \cite{LE18,LL20},
and (ix) \textbf{pressures} \cite{LE18,LL20}. The generalized Debye-H\"{u}ckel
formula (\ref{6.15}) includes all these 9 physical properties with only 3
fitting parameters. However, we look forward to the day when we can derive the
three fitting parameters for particular types of ions, from independently
determined experimental data.

\subsection{Electric Double Layers}

We consider a charged surface in contact with a 0.1 M 1:4 aqueous electrolyte,
where the charge density is $\sigma=1e/(50$\AA $^{2})$, the radius of both
cations and anions is $a=4.65$ \AA \ (in contrast to an edge length of 7.5
\AA \ of cubical ions in \cite{BA97}), and $\epsilon_{s}=80$ \cite{LX17}. The
multivalent ions represent large polyanions adsorbed onto a charged Langmuir
monolayer in experiments \cite{BA97}. We solve (\ref{6.3}) and (\ref{6.4})
using (\ref{7.1}) in the rectangular box $\overline{\Omega}=\overline{\Omega
}_{s}=\left\{  (x,y,z):0\leq x\leq40\text{, }-5\leq y\leq5\text{, }-5\leq
z\leq5\text{ \AA }\right\}  $ such that $\phi(\mathbf{r})\approx0$ within the
accuracy to $10^{-4}$ near and on the surface $x=40$ \AA . The boundary
conditions on the surface and its adjacent four planes are $-\epsilon
_{s}\nabla\phi\cdot\mathbf{n}=\sigma$ with $\mathbf{n}=\left\langle
-1,0,0\right\rangle $ and $-\epsilon_{s}\nabla\phi\cdot\mathbf{n}=0$ with
$\mathbf{n}$ defined similarly, respectively.

The classical PB model (with $a=a_{\text{H}_{2}\text{O}}=l_{c}=0$, i.e., no
size, void, and correlation effects) produces unphysically high concentrations
of anions (A$^{4-}$) near the surface as shown by the dashed curve in Fig. 6.
The dotted curve in Fig. 6 is similar to that of the modified PB in
\cite{BA97} and is obtained by the 4PBik model with $l_{c}=0$ (no
correlations), $V_{K+2}=0$ (no voids), and $a_{\text{H}_{2}\text{O}}=0$ (water
is volumeless as in \cite{BA97} and hence $\Gamma^{B}=1-\sum_{i=1}^{K}%
v_{i}C_{i}^{B}$ is the bulk water volume fraction). The voids ($V_{K+2}\neq0$)
and water molecules ($a_{\text{H}_{2}\text{O}}\neq0$) have slight effects on
anion concentration (because of saturation) and electric potential (because
water and voids have no charges) profiles as shown by the thin solid curves in
Figs. 6 and 7, respectively, when compared with the dotted curves. However,
ion-ion correlations (with $l_{c}=1.6a$ \cite{BS11}) have significant effects
on ion distributions as shown by the thick solid and dash-dotted curves in
Fig. 6, where the saturation layer is on the order of ionic radius $a$ and the
\textbf{overscreening} layer \cite{BS11} ($C_{\text{A}^{4-}}(x)\approx
0<C_{\text{A}^{4-}}^{B}=0.1$ M) of excess coions ($C_{\text{C}^{+}}(x)$
$>$
$C_{\text{C}^{+}}^{B}=0.4$ M) is about 18 \r{A}\ in thickness.

The \textbf{saturation layer} is an \textbf{output} (not an imposed condition)
of our model unlike a Stern layer\ \cite{S24} imposed by most EDL models to
account for size effects near charge surfaces \cite{O08,GR11,BG16}. The
electric potentials $\phi(0)=$ 5.6 at $x=0$ and $\phi(11.5)=$ -1.97 $k_{B}T/e$
in Fig. 7 obtained by 4PBik with voids and correlations \textbf{deviate
}dramatically from those by previous models for nearly \textbf{100\%} at $x=0$
(in the saturation layer) and \textbf{70\%} at $x=11.5$ \r{A}\ (in the
screening layer) when compared with the maximum potential $\phi(0)=2.82$
$k_{B}T/e$ of previous models. The 4PBik potential depth $\phi(11.5)=-1.97$
$k_{B}T/e$ of the overscreening layer is very sensitive the size $a$ of ions
and tends to zero as $a\rightarrow0$.\begin{figure}[t]
\centering\includegraphics[scale=0.6]{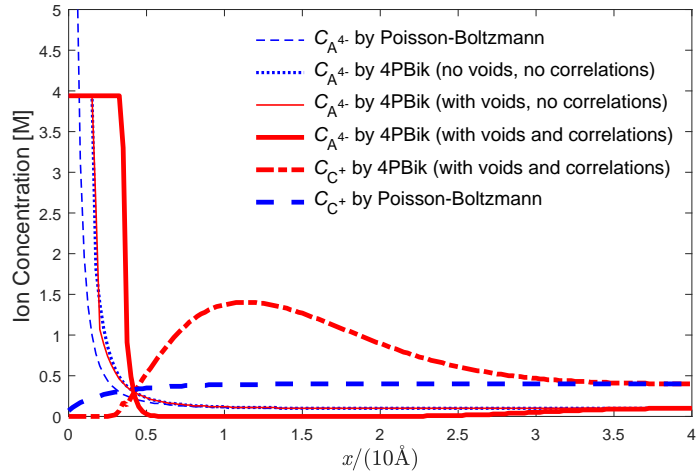}\caption{Concentration
profiles of anions $C_{\text{A}^{4-}}(x)$ and cations $C_{\text{C}^{+}}(x)$
obtained by various models in a C$_{4}$A electrolyte solution with the charge
density $\sigma=1e/(50$\r{A}$^{2})$ at $x=0$.}%
\end{figure}\begin{figure}[tt]
\centering\includegraphics[scale=0.6]{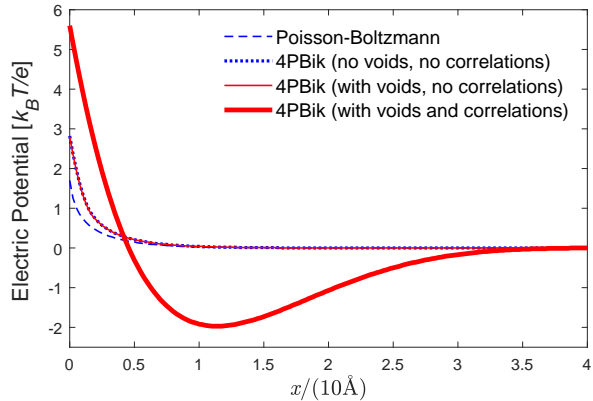}\caption{Electric potential
profiles $\phi(x)$.}%
\end{figure}

\subsection{Biological Ion Channels}

Biological ion channels are a particularly appropriate test of a model of
concentrated ionic solutions.

The data available for tens to hundreds of different types of channels and
transporters is breathtaking: it is often accurate to a few per cent (because
signal to noise ratios are so large and biological variation hardly exists for
channels of known amino acid sequence, which means nearly every channel
nowadays). The data is always nonequilibrium, i.e., current voltage relations
in a wide range of solutions of different composition and concentration, or
(limiting zero voltage) conductance in those solutions. Indeed, many of the
channels do not function if concentrations are equal on both sides and the
electrical potential is zero. They are said to inactivate.

The data is often available for single channels recorded individually in patch
clamp or bilayer configuration. Data is available for a range of divalent
(usually calcium ion) concentrations because calcium concentration is often a
controller of channel, transporter, and biological activity in the same sense
that a gas pedal is the controller of the speed of a car. The structure of the
ion channel or transporter is often known in breathtaking detail. The word
`breathtaking' is appropriate because similar structures are rarely if ever
known of strictly physical systems. The structure and the structure of the
permanent and polarization charge of the channel protein (that forms the pore
through which ions move) can be modified by standard methods of site directed
mutagenesis, for example, that are available in `kit' form usable by most
molecular biology laboratories. Thus, models can be tested from atomic detail
to single channel function to ensemble function to cellular and physiological
function, even to the ultimate biological function (like the rate of the heart
beat). Few other systems allow experimental measurement at each level of the
hierarchy linking the atomic composition of genes (that encode the channel's
amino acid composition), to the atomic structure of the channel, right to the
function of the cell. The hierarchy here reaches from 10$^{-11}$ to 10$^{-5}$
meters. When the channel controls the biological function of an organ like the
heart, the hierarchy reaches to $2\cdot10^{-1}$ meters, in humans for example.

The biological significance of ion channels is hard to exaggerate since they
play a role in organisms analogous to the role of transistors in computers.
They are the device that execute most of the physical controls of current and
ion movement that are then combined in a hierarchy of structures to make
biological cells, tissues, and organisms, if not populations of organisms.

From a physical point of view, ion channels provide a particularly crowded
environment in which the effects of the steric potential (crowding in more
traditional language) and electrical potential can combine to produce striking
characteristics of selectivity and rectification. Theories that do not deal
explicitly with ion channel data, i.e., that do not predict current voltage
relations from known structures, seem to us to be begging central PHYSICAL
questions that might falsify their approach. In fact, as a matter of history
it is a fact that we learned how to construct our model of bulk solutions from
our earlier work on ion channels.

\subsubsection{Gramicidin A Channel}

We use the gramicidin A (GA) channel in Fig. 2 to illustrate the full
Poisson-Nernst-Planck-Bikerman theory consisting of Eqs. (\ref{2.4}),
(\ref{5.1}), (\ref{5.2}), (\ref{6.2}), (\ref{6.3}), (\ref{6.4}), and
conditions (\ref{6.5}) - (\ref{6.9}) with --- steric, correlation,
polarization, dielectric, charge/space competition, and nonequilibrium effects
--- at steady state using the algorithm and methods in Section 3 to perform
numerical simulations. The union domain $\overline{\Omega}_{i}\cup\Omega_{sh}$
in Fig. 1 is replaced by the biomolecular domain $\Omega_{m}$ in Fig. 3.

Fig. 8 shows I-V curves obtained by PNPB and compared with experimental data
(symbols) by Cole et al. \cite{CF02} with bath K$^{+}$ and Cl$^{-}$
concentrations $C^{B}=0.1$, 0.2, 0.5, 1, 2 M and membrane potentials $\Delta
V=0$, 50, 100, 150, 200 mV. The PNPB currents in pico ampere (pA) were
obtained with $\theta=1/4.7$ in the pore diffusion coefficients $\theta D_{i}$
from (\ref{5.4}) for all particle species. The reduction parameter $\theta$
has been used in all previous PNP papers and is necessary for continuum
results to be comparable to MD, BD, or experimental data \cite{G08}. The
values of other model parameters are listed in Table I in \cite{LE15}.
\begin{figure}[t]
\centering\includegraphics[scale=0.6]{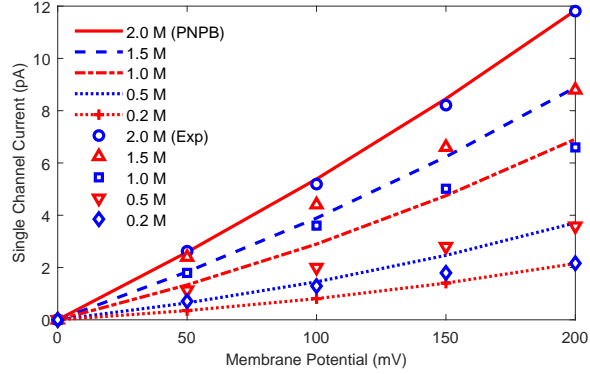}\caption{Comparison of PNPB
(curves) and experimental \cite{CF02} (symbols) I-V results with bath K$^{+}$
and Cl$^{-}$ concentrations $C^{\text{B}}=0.1$, 0.2, 0.5, 1, 2 M and membrane
potentials $\Delta V=0$, 50, 100, 150, 200 mV.}%
\end{figure}\begin{figure}[tt]
\centering\includegraphics[scale=0.6]{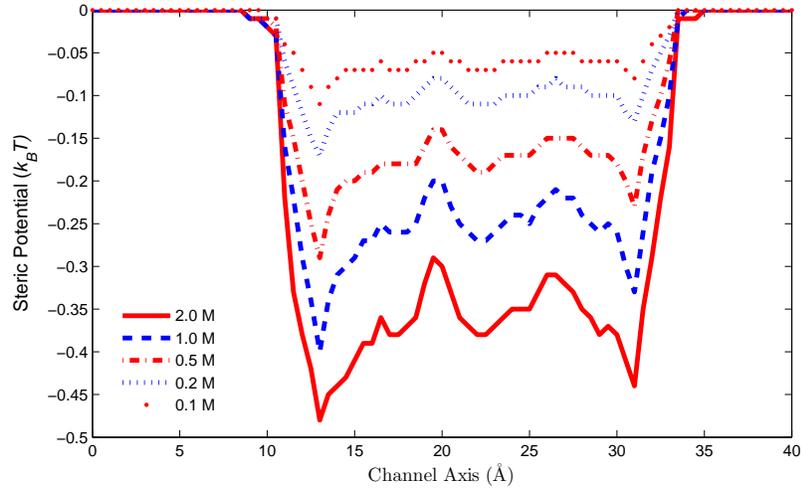}\caption{Averaged steric
potential $S^{trc}(\mathbf{r})$ profiles at each cross section along the pore
axis with $C^{B}=$ 0.1, 0.2, 0.5, 1, 2 M and $\Delta V=200$ mV. The same
averaging method applies to the following profiles.}%
\end{figure} \begin{figure}[ttt]
\centering\includegraphics[scale=0.6]{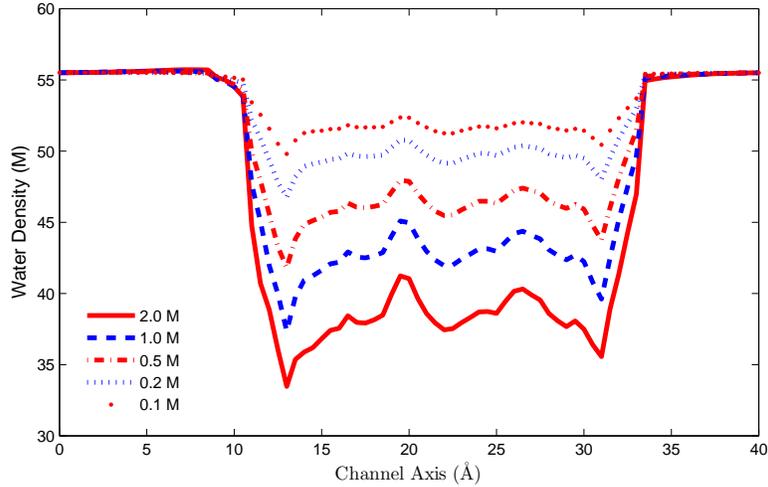}\caption{Water density
$C_{\text{H}_{2}\text{O}}(\mathbf{r})$ profiles.}%
\end{figure}\begin{figure}[tttt]
\centering\includegraphics[scale=0.6]{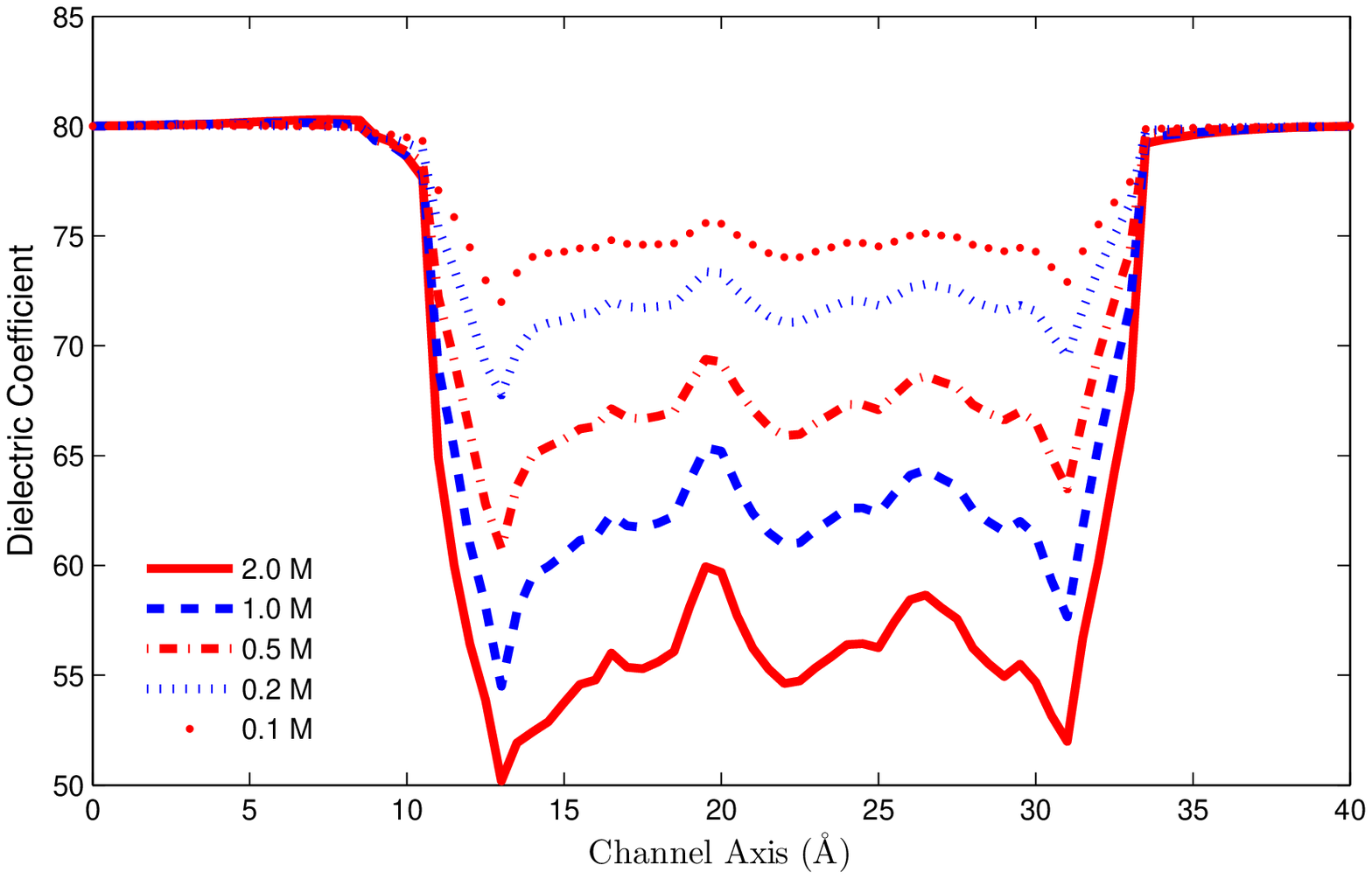}\caption{Dielectric function
$\widetilde{\epsilon}(\mathbf{r})$ profiles.}%
\end{figure}

We summarize the novel results of PNPB in \cite{LE15} when compared with those
of earlier continuum models for ion channels: (i) The pore \textbf{diffusion}
parameter $\theta=1/4.7$ agrees with the range 1/3 to 1/10 obtained by many MD
simulations of various channel models \cite{SS98,AK00,MC03} indicating that
the steric (Fig. 9), correlation, dehydration (Fig. 10), and dielectric (Fig.
11) properties have made PNPB simulations more closer (realistic) to MD than
previous PNP for which $\theta$ differs from MD values by an order to several
orders of magnitude \cite{AK00}. (ii) Figs. 9, 10, and 11, which are all
absent in earlier work, show that these properties \textbf{correlate} to each
other and \textbf{vary} with salt concentration and protein charges in a
\textbf{self-consistent} way by PNPB. (iii) The steric potential profiles in
Fig. 9 clearly illustrate the \textbf{charge/space competition} between ions
and water under dynamic and variable conditions. For example, the global
minimum value in Fig. 9 at $\widehat{r}=13.1$ on the channel axis, where the
channel protein is most negatively charged, is $S^{trc}(\widehat{r})=\ln
\frac{\Gamma(\widehat{r}\mathbf{)}}{\Gamma^{B}}=-0.485$ yielding
$\Gamma(\widehat{r}\mathbf{)/}\Gamma^{B}=0.616$. Namely, it is 38.4\% more
crowded at $\widehat{r}$ than in the bath and mainly occupied by K$^{+}$ as
shown in Figs. 10 and 12. It is important to \textbf{quantify voids}
($\Gamma(\mathbf{r)}=1-\sum_{i=1}^{K+1}v_{i}C_{i}(\mathbf{r})$) at highly
charged locations in channel proteins and many more biological, chemical, and
nano systems. The charge space competition has been a central topic in the
study of ion channels since at least \cite{CN95,NE98,NC00,BB02,E03}. The
literature is too large to describe in detail here. Recent reviews can help
\cite{BH13,B14,G15,MV17}. (iv) PNPB preserves \textbf{mass conservation} due
to void and size effects in contrast to PNP as shown in Fig. 13, where the
total number of H$_{2}$O and K$^{+}$ in the channel pore is 8 \cite{RP95}%
.\begin{figure}[t]
\centering\includegraphics[scale=0.6]{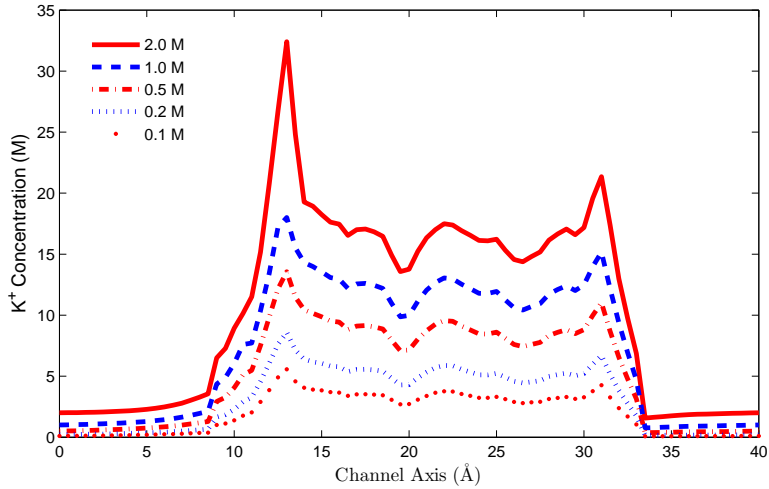}\caption{K$^{+}$ concentration
$C_{\text{K}^{+}}(\mathbf{r})$ profiles.}%
\end{figure}\begin{figure}[tt]
\centering\includegraphics[scale=0.6]{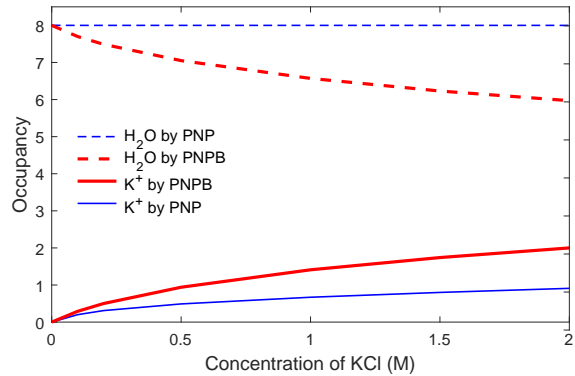}\caption{Occupancy of H$_{2}$O
and K$^{+}$ in the GA channel pore by PNPB and PNP as [KCl] increases from 0
to 2 M. The total number of H$_{2}$O and K$^{+}$ in the pore is 8 \cite{RP95},
which is conserved by PNPB but not by PNP (without steric and correlation
effects).}%
\end{figure}

\subsubsection{L-type Calcium Channel}

L-type calcium channels operate very delicately in physiological and
experimental conditions. They exquisitely tune their conductance from
\textbf{Na}$^{+}$\textbf{-flow}, to \textbf{Na}$^{+}$\textbf{-blockage}, and
to \textbf{Ca}$^{2+}$\textbf{-flow} when bath Ca$^{2+}$ varies from trace to
high concentrations as shown by the single-channel currents in femto ampere in
Fig. 14 (circle symbols) recorded by Almers and McCleskey \cite{AM84}, where
the range of extracellular concentrations [Ca$^{2+}$]$_{\text{o}}$ is
\textbf{10}$^{8}$\textbf{-fold} from $10^{-10}$ to $10^{-2}$
M.\begin{figure}[t]
\centering\includegraphics[scale=0.6]{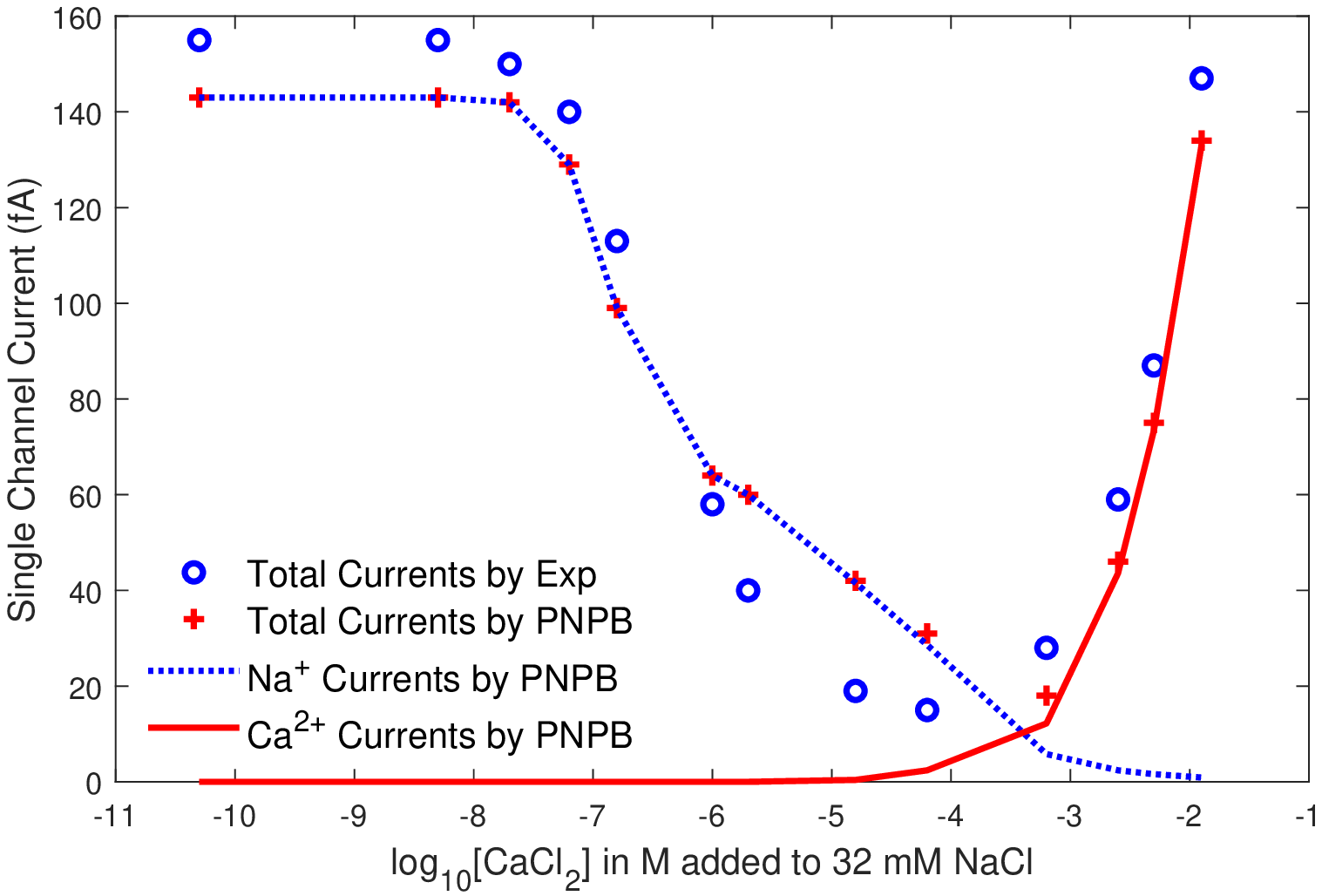}\caption{Single channel
currents in femto ampere (fA) plotted as a function of $\log_{10}$[Ca$^{2+}%
$]$_{\text{o}}$. Experimental data of \cite{AM84} are marked by small circles
and PNPB data are denoted by the plus sign and lines.}%
\end{figure}\begin{figure}[tt]
\centering\includegraphics[scale=1.0]{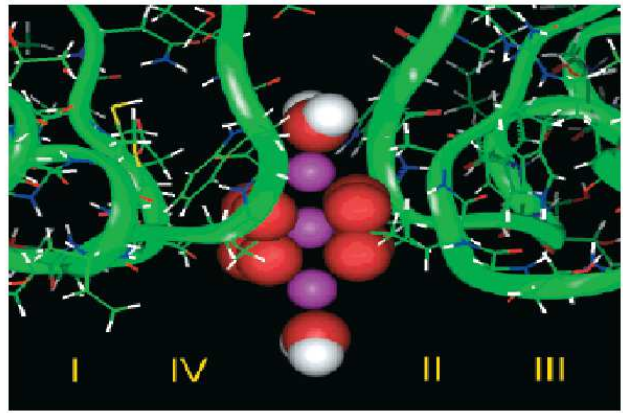}\caption{The Lipkind-Fozzard
pore model of L-type calcium channel, where 3 Ca$^{2+}$ are shown in violet, 8
O$^{1/2-}$ in red, 2 H$_{2}$O in white and red. Reprinted with permission from
(G. M. Lipkind and H. A. Fozzard, Biochem. \textbf{40}, 6786 (2001)).
Copyright (2001) American Chemical Society.}%
\end{figure}

We used the Lipkind-Fozzard molecular model \cite{LF01} shown in Fig. 15 to
perform PNPB simulations with both atomic and continuum methods (Algorithm 2
in \cite{LE15}) for this model channel, where the EEEE locus (four glutamate
side chains modeled by 8 O$^{1/2-}$ ions shown by red spheres) forms a
high-affinity Ca$^{2+}$ binding site (center violet sphere) that is essential
to Ca$^{2+}$ selectivity, blockage, and permeation. Water molecules are shown
in white and red. More realistic structures would be appropriate if the work
were done now, but the analysis here shows the ability of PNPB to deal with
experimental data using even a quite primitive model of the structure.

PNPB results (plus symbols) in Fig. 14 agree with the experimental data at
[$\text{Na}^{+}$]$_{\text{i}}=$ [$\text{Na}^{+}$]$_{\text{o}}=32$ mM,
[Ca$^{2+}$]$_{\text{i}}=0$, $V_{\text{o}}=0$, and $V_{\text{i}}$ $=-20$ mV
(the intracellular membrane potential), where the partial Ca$^{2+}$ and
Na$^{+}$ currents are denoted by the solid and dotted line, respectively.
These two ionic currents show the anomalous fraction effect of the channel at nonequilibrium.

\subsubsection{Potassium Channel}

Potassium channels conduct K$^{+}$ ions very rapidly (nearly at the diffusion
rate limit (10$^{8}$ per second) in bulk water) and \textbf{selectively}
(excluding, most notably, Na$^{+}$ despite their difference in radius is only
$a_{\text{K}^{+}}-a_{\text{Na}^{+}}=1.33-0.95=0.38$ \r{A}\ in
\textbf{sub-Angstrom} range) \cite{H01}. Fig. 16 shows the structure of KcsA
(PDB ID 3F5W) crystallized by Cuello et al. \cite{CJ10}, where the spheres
denote five specific cation binding sites (S0 to S4) \cite{NB04} in the
solvent domain $\Omega_{s}$ and the channel protein in $\Omega_{m}$ consists
of $N=31268$ charged atoms. The exquisite selectivity of K$^{+}$ over Na$^{+}$
by K channels can be quantified by the free energy ($G$) differences of
K$^{+}$ and Na$^{+}$ in the pore and in the bulk solution, i.e., by $\Delta
G($K$^{+})=\left[  G_{\text{pore}}(\text{K}^{+})-G_{\text{bulk}}(\text{K}%
^{+})\right]  $ and $\Delta G($Na$^{+})=G_{\text{pore}}($Na$^{+}%
)-G_{\text{bulk}}($Na$^{+})$ \cite{NB04}. Experimental measurements
\cite{NM88,LH01,NM02} showed that the relative free energy
\begin{equation}
\Delta\Delta G(\text{K}^{+}\rightarrow\text{Na}^{+})=\Delta G(\text{Na}%
^{+})-\Delta G(\text{K}^{+})=5\sim6\text{ kcal/mol}\label{8.1}%
\end{equation}
unfavorable for Na$^{+}$. \begin{figure}[t]
\centering\includegraphics[scale=0.8]{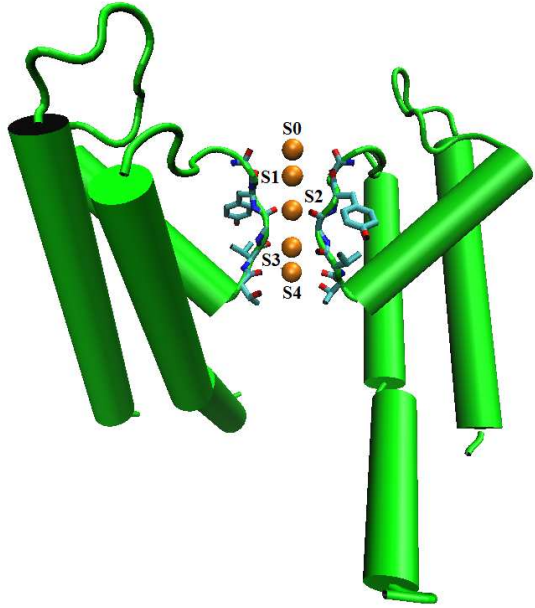}\caption{The crystal structure
of the K channel KcsA (PDB ID 3F5W) \cite{CJ10} with five cation binding sites
S0, S1, S2, S3, and S4 \cite{NB04} marked by spheres.}%
\end{figure}

Free energies can be calculated by the electric and steric potentials
\cite{LX17}%
\begin{equation}
\phi_{\text{S2}}=\frac{1}{4\pi\epsilon_{0}}\left(  \frac{1}{6}\sum_{k=1}%
^{6}\sum_{j=1}^{N}\frac{q_{j}}{\epsilon_{p}(r)|c_{j}-A_{k}|}+\frac
{q_{\text{S2}}}{\epsilon_{b}a_{\text{S2}}}\right)  \text{, }S_{\text{S2}%
}^{trc}=\ln\frac{1-\frac{v_{\text{S2}}}{V_{\text{S2}}}}{\Gamma^{B}},
\label{8.2}%
\end{equation}
at the binding site S2 \cite{NB04} on the atomic scale, where S2 also denotes
Na$^{+}$ or K$^{+}$ (the site is occupied by a Na$^{+}$ or K$^{+}$), $q_{j}$
is the charge on the atom $j$ in the protein given by PDB2PQR \cite{DC07},
$\epsilon_{p}(r)=1+77r/(27.7+r)$ \cite{MM02}, $r=|c_{j}-$ $c_{\text{S2}}|$,
$c_{j}$ is the center of atom $j$, $A_{k}$ is one of six symmetric surface
points on the spherical S2, $\epsilon_{b}=3.6$, and $V_{\text{S2}%
}=1.5v_{\text{K}^{+}}$ is a volume containing the ion at S2. We obtained
$\Delta\Delta G=5.26$ kcal/mol \cite{LX17} in accord with the MD result $5.3$
kcal/mol \cite{NB04}, where $G_{\text{pore}}($Na$^{+})=4.4$, $G_{\text{bulk}%
}($Na$^{+})=-0.26$, $G_{\text{pore}}($K$^{+})=-0.87$, $G_{\text{bulk}}($%
K$^{+})=-0.27$ kcal/mol, $\phi_{\text{Na}^{+}}=7.5$ $k_{B}T/e$, $\frac
{v_{\text{Na}^{+}}}{v_{0}}S_{\text{Na}^{+}}^{trc}=0.23$, $\phi_{\text{K}^{+}%
}=-1.93$ $k_{B}T/e$, $\frac{v_{\text{K}^{+}}}{v_{0}}S_{\text{K}^{+}}%
^{trc}=-0.59$, and $C_{\text{Na}^{+}}^{B}=C_{\text{K}^{+}}^{B}=0.4$ M.

The \textbf{crucial} parameter in (\ref{8.2}) is the \textbf{ionic radius}
$a_{\text{S2}}=0.95$ or $1.33$ \AA \ (also in $|c_{j}-A_{k}|$) that affects
$\phi_{\text{S2}}$ very strongly but $S_{\text{S2}}^{trc}$ weakly. Another
\textbf{important} parameter in (\ref{8.2}) is the \textbf{bulk void fraction}
$\Gamma^{B}$ that depends on the bulk concentrations of all ions and water and
links the total energy of the ion at S2 to these bulk conditions measured
\textbf{very far} away ($\sim$ 10$^{6}$ \AA ) in the baths on the atomic scale.

\subsubsection{Sodium Calcium Exchanger}

The Na$^{+}$/Ca$^{2+}$ exchanger (NCX) is the major cardiac mechanism that
\textbf{extrudes} intracellular Ca$^{2+}$ across the cell membrane
\textbf{against} its chemical gradient by using the \textbf{downhill} gradient
of Na$^{+}$ \cite{SO15}. The molecular basis of Na$^{+}$/Ca$^{2+}$
interactions in NCX so striking to L\"{u}ttgau and Niedegerke \cite{LN58} have
been revealed by the cloning of NCX gene \cite{NL90} and the structure of the
ancient archaebacterial version NCX\_Mj determined by Liao et al. \cite{LL12}.
Fig. 17 illustrates NCX\_Mj viewed from the membrane, which consists of 10
transmembrane helices that form a binding pocket of three putative Na$^{+}$
(green spheres) and one Ca$^{2+}$ (blue sphere) binding sites \cite{LL12}%
.\begin{figure}[t]
\centering\includegraphics[scale=1.0]{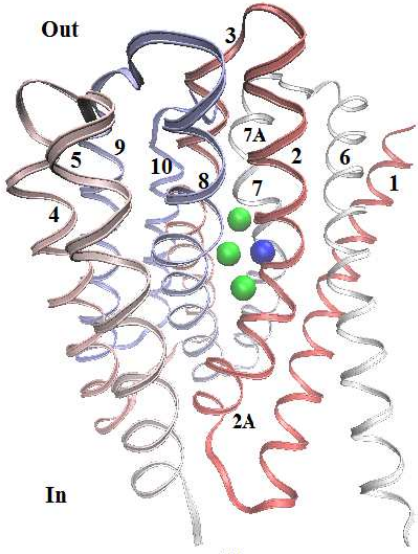}\caption{Structure of NCX\_Mj
consisting of ten transmembrane helices that form a binding pocket of three
Na$^{+}$ (green spheres) and one Ca$^{2+}$ (blue sphere) binding sites
\cite{LL12}.}%
\end{figure}\begin{figure}[tt]
\centering\includegraphics[scale=1.0]{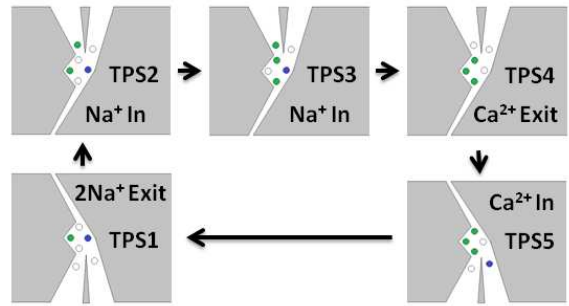}\caption{Schematic diagram of
a cycle of Na$^{+}$/Ca$^{2+}$ exchange in NCX consisting of five total
potential states (TPS). Two Na$^{+}$ and one Ca$^{2+}$ ions enter the binding
pocket in the outward- (TPS2 $\rightarrow$ TPS3 $\rightarrow$ TPS4) and
inward-facing (TPS5 $\rightarrow$ TPS1) conformations, respectively. They exit
in opposite conformations \cite{LH16}.}%
\end{figure}

We developed a cyclic model of Na$^{+}$/Ca$^{2+}$ exchange mechanism in NCX
\cite{LH16} using (\ref{8.2}) to calculate five total (electric and steric)
potential states (TPS) of various Na$^{+}$ and Ca$^{2+}$ ions shown in Fig.
18, where TPS1 and TPS4 are stable (having negative values) and TPS2, TPS3,
and TPS5 are unstable (positive). Four extra sites in Fig. 18 are determined
empirically and close to entrance or exit locations of the binding pocket. The
green and blue dots in the diagram represent Na$^{+}$ and Ca$^{2+}$ ions
occupying the respective sites. Two Na$^{+}$ and one Ca$^{2+}$ ions enter the
binding pocket in the outward- (TPS2 $\rightarrow$ TPS3 $\rightarrow$ TPS4)
and inward-facing (TPS5 $\rightarrow$ TPS1) conformations, respectively. They
exit in opposite conformations. The cycle consists of five steps.

\emph{Step 1:} A conformational change is hypothetically activated \cite{LH16}
by a binding Ca$^{2+}$ at the blue site (S1) in TPS1 from inward-facing to
outward-facing in TPS2.

\emph{Step 2:} One Na$^{+}$ enters the binding pocket from the access site in
TPS2 to the top Na$^{+}$ binding site (S2) in TPS3 followed by another
Na$^{+}$ to the access site. These two coming Na$^{+}$ ions move the existing
Na$^{+}$ ion from the middle Na$^{+}$ site (S3) to the bottom site (S4) by
their Coulomb forces. TPS2 and TPS3 are unstable meaning that the two coming
Na$^{+}$ ions have positive energies and are thus mobile. The
\textbf{selectivity} ratio of Na$^{+}$ to Ca$^{2+}$ by NCX from the
extracellular bath to the binding site S2 is $C_{\text{Na}^{+}}($%
S2$)/C_{\text{Ca}^{2+}}(S2)=55.4$ under the experimental conditions of the
extracellular bath $\left[  \text{Na}^{+}\right]  _{\text{o}}=120$ mM and
$\left[  \text{Ca}^{2+}\right]  _{\text{o}}=1$ $\mu$M \cite{LH16}.

\emph{Step 3:} The vacant site S3 in TPS3 is a deep potential well having TP =
$-8.89$ $k_{B}T/e$ that pulls the two unstable Na$^{+}$ ions to their sites in
TPS4. Meanwhile, these two moving Na$^{+}$ and the stable Na$^{+}$ at S4
extrude the Ca$^{2+}$ (having unstable TP = 1.65) at S1 out of the pocket to
become TPS4.

\emph{Step 4:} Now, all three Na$^{+}$ ions in TPS4 are stable having negative
TP and the vacant site S1 has an even deeper TP = $-16.02$. We conjecture that
this TP value may trigger a conformational change from outward-facing in TPS4
to inward-facing in TPS5. The mechanism of conformational changes in NCX is
yet to be studied.

\emph{Step 5:} Furthermore, this large negative TP in TPS5 yields a remarkably
large \textbf{selectivity} ratio of Ca$^{2+}$ to Na$^{+}$ by NCX from the
intracellular bath to S1, i.e., $C_{\text{Ca}^{2+}}($S1$)/C_{\text{Na}^{+}}%
($S1$)=4986.1$ at $\left[  \text{Ca}^{2+}\right]  _{\text{i}}=33$ $\mu$M and
$\left[  \text{Na}^{+}\right]  _{\text{i}}=60$ mM. A coming Ca$^{2+}$ in TPS5
then extrudes two Na$^{+}$ ions out of the packet when it settles at S1 in
stable TPS1.

Assuming that the total time T of an exchange cycle is equally shared by the 5
TPS, this model also infers that the \textbf{stoichiometry} of NCX is
$\frac{3}{5}$T$\cdot2$ Na$^{+}:\frac{2}{5}$T$\cdot1$ Ca$^{2+}=3$ Na$^{+}:1$
Ca$^{2+}$ in transporting Na$^{+}$ and Ca$^{2+}$ ions \cite{LH16}, which is
the generally accepted stoichiometry (see reviews of Blaustein and Lederer
\cite{BL99a} and Dipolo and Beaug\'{e} \cite{DB06}) since the pivotal work of
Reeves and Hale \cite{RH84} and other subsequent experimental results.

\section{Discussion and Conclusions}

We have covered a range of aspects of the fourth-order
Poisson-Nernst-Planck-Bikerman theory from physical modeling, mathematical
analysis, numerical implementation, to applications and verifications for
aqueous electrolyte systems in chemistry and biology. The theory can describe
many properties of ions and water in the system that classical theories fail
to describe such as steric, correlation, polarization, variable permittivity,
dehydration, mass conservation, charge/space competition, overscreening,
selectivity, saturation, and more. All these properties are accounted for in a
single framework with only two fundamental parameters, namely, the dielectric
constant of pure water and the correlation length of empirical choice. Ions
and water have their physical volumes as those in molecular dynamic
simulations. The theory applies to a system at both continuum and atomic
scales due to the exact definition of the total volume of all ions, water
molecules, and interstitial voids.

The most important features of PNPB are that (i) ions and water have unequal
volumes with interstitial voids, (ii) their distributions are saturating of
the Fermi type, (iii) these Fermi distributions approach Boltzmann
distributions as the volumes tend to zero, and (iv) all the above physical
properties appear self-consistently in a single model not separately by
various models. Most existing modified Poisson-Boltzmann models consider ions
of equal size and fail to yield Boltzmann distributions in limiting cases,
i.e., the limit is divergent indicating that steric energies are poorly
estimated. Numerous models for different properties such as steric,
correlation, polarization, permittivity are proposed separately in the past.

We have shown how to solve 4PBik analytically and PNPB numerically. The
generalized Debye-H\"{u}ckel theory derived from the 4PBik model gives
valuable insights into physical properties and leads to an electrolyte
(analytical) equation of state that is useful to study thermodynamic
activities of ion and water under wide ranges of composition, concentration,
temperature, and pressure.

Numerically solving the fourth-order PNPB model in 3D for realistic problems
is a challenging task. There are many pitfalls that one must carefully avoid
in coding. For that reason, we have particularly mentioned some methods for
handling the convergence issues of the highly nonlinear PNPB system of partial
differential equations and the discretization problems concerning the
complicate interface between molecular and solvent domains and the
Scharfetter-Gummel stability condition to ensure positivity of numerical
concentrations and current preservation.

Finally, we have shown novel results obtained by PNPB for chemical and
biological systems on ion activities, electric double layers, gramicidin A
channel, L-type calcium channel, potassium channel, and sodium calcium
exchanger. These results agree with experiments or molecular dynamics data and
show not only continuum but also atomic properties of the system under far
field conditions. The fourth-order PNPB model is consistent and applicable to
a great variety of systems on a vast scale from \textbf{meter} to
\textbf{Angstrom}.

\end{document}